\documentclass[letterpaper,journal]{IEEEtran}

\usepackage{amsmath,amsfonts,amssymb}
\usepackage{algorithm}
\usepackage{algpseudocode}
\usepackage{booktabs}
\usepackage{multirow}
\usepackage{array}
\usepackage{graphicx}
\usepackage{url}
\usepackage{cite}
\usepackage{balance}
\usepackage{xcolor}

\newcommand{\safeincludegraphics}[2][]{%
\IfFileExists{#2}{\includegraphics[#1]{#2}}{\fbox{\parbox{0.92\linewidth}{Missing figure: {\ttfamily\detokenize{#2}}}}}%
}

\begin{document}
\bstctlcite{BSTcontrol}
\title{SCORAS-MoE: Joint Compression and Resource-Adaptive Deployment of MoE-VLMs in LEO Satellite Networks}

\author{Tong~Quan, ~Yuanlong~Wan, ~Huasen~He,~\IEEEmembership{Member,~IEEE}, ~Yunpeng~Hou, ~Shuangwu~Chen, ~Xiaofeng~Jiang, ~Jian~Yang,~\IEEEmembership{Senior Member,~IEEE}
    \thanks{T. Quan, Y. Wan, H. He, Y. Hou, S. Chen, X. Jiang, and J. Yang are with the Department of Automation, University of Science and Technology of China, Hefei 230027, China, and also with the Institute of Artificial Intelligence, Hefei Comprehensive National Science Center, Hefei 230026, China.}
 \thanks{H. He (hehuasen@ustc.edu.cn) is the corresponding author.}
}

\markboth{Preprint}%
{Quan et al.: SCORAS-MoE}

\maketitle

\begin{abstract}
Deploying large vision--language models (VLMs) onboard satellites
enables onboard data processing and reduces raw data downlink. However, onboard inference faces two resource challenges.
Limited onboard memory and energy require model compression
and distributed deployment. Dynamic resource availability requires
fast deployment decisions as illumination, battery levels,
and communication conditions change.
We present \textbf{SCORAS-MoE},
a joint compression and deployment framework for mixture-of-experts (MoE)
VLMs in low Earth orbit (LEO) satellite networks. To address limited resources, SCORAS-MoE
measures the perturbation of the routed MoE output caused by low-rank
approximation, assigns higher ranks to more sensitive experts, and distributes compressed model shards
across satellites for cooperative inference. The compressed models yield profiles of measured accuracy
and inference energy. To adapt to dynamic resources, the online
scheduler selects profile compositions and shard placements in each
slot. For each candidate composition, it reduces placement to a
minimum-cost assignment problem solved by the Hungarian algorithm, while
enumerating the compositions yields the optimal deployment for the
current-slot objective. Experiments on Qwen3-VL-30B-A3B-Instruct show
that allocating ranks based on output perturbation is particularly
effective under aggressive compression, with an absolute gain of
$3.7\%$ in mean accuracy over uniform rank allocation when expert
projections retain $30\%$ of their original parameters. The fixed-profile scheduler
achieves higher throughput with fewer service switches and lower
battery impact than the evaluated proximal policy optimization (PPO)
and evolutionary baselines, with respective speedups of $8.7\times$
and $183.5\times$.
Adaptive profile selection further improves the balance between
service quality and energy use.
\end{abstract}


\section{Introduction}
\label{sec:introduction}

\IEEEPARstart{S}{atellites} continuously collect imagery and other
multimodal observations. Large vision--language models (VLMs) can turn
these observations into descriptions and answers directly onboard, reducing the need to transmit raw data through limited
downlink opportunities. Processing observations where they are generated
also supports timely services when ground stations are unavailable
\cite{li2026grace}. The NAVI-Orbital demonstration illustrates this
potential: a Gemma~3 VLM deployed on a spacecraft in low Earth orbit (LEO) classified and
described newly captured Earth imagery and supported natural-language
interaction onboard~\cite{delfa2026navi}. Bringing these capabilities
to sustained satellite services requires addressing two resource
challenges: resources are both \emph{limited} and \emph{dynamic}.

\emph{Limited resources require effective compression and distributed deployment.}
Satellites have constrained memory, computing capacity, and energy
supply, making large VLMs difficult to deploy. Mixture-of-experts (MoE)
architectures, including those in the Qwen3-VL family~\cite{bai2025qwen3vl},
offer high model capacity while activating only a subset of experts
for each token. The complete expert set nevertheless retains a large
parameter count. Low-rank compression reduces the model's memory and
computation requirements, but achieving a small target parameter count
requires careful rank allocation: experts
differ in how strongly their approximation affects the model output.
The central compression question is therefore how to distribute
a fixed total rank across experts while preserving inference quality. Even after
compression, the model may exceed the capacity of a single satellite,
requiring its shards to be distributed across multiple satellites for
cooperative inference.

\emph{Dynamic resources require fast deployment decisions.}
Orbital motion changes solar-energy availability and communication
conditions, while inference continuously affects battery states.
A deployment suited to one slot may be inefficient or infeasible in
the next. Within this distributed deployment, the service must therefore
quickly adjust both the selected model profiles and shard placement,
while accounting for the cost of reconfiguration.

Existing research provides foundations for both challenges. MoE
compression exploits expert heterogeneity through pruning, quantization,
and low-rank approximation~\cite{chen2025eacmoe,xie2025automated,qi2026profiling}.
Structural and routing statistics offer inexpensive allocation signals,
while output-space distortion provides a functional measure of
compression sensitivity~\cite{lu2024experts,duanmu2025mxmoe,deng2026gemq}.
Satellite edge systems study adaptive deployment, model partitioning,
offloading, and energy-aware execution
\cite{yao2025leoedge,fan2025satellite,chen2025slice,
liu2024phoenix,shi2025satellite_lam}. Connecting compression to online
deployment remains essential: the compression choice determines
inference quality and resource demand, and the deployment determines
whether that choice can serve requests under the current satellite
state. An effective framework needs both accurate compressed models
and a fast way to place them as resources change.

We propose \textbf{SCORAS-MoE}
(\textbf{S}ensitivity-Guided \textbf{Co}mpression and
\textbf{R}esource-\textbf{A}daptive \textbf{S}cheduling for MoE),
a joint compression and deployment framework for LEO satellite
networks. Offline, it measures the perturbation of the routed MoE
output caused by low-rank approximation and assigns higher ranks to
more sensitive experts.
Compressing and evaluating the model at different parameter retention ratios produces
a set of accuracy--energy profiles. Online, it uses these profiles
to adapt deployment to the current resource state. For each candidate
profile composition, shard placement becomes a minimum-cost assignment
problem solved by the Hungarian algorithm. Comparing the candidate
compositions gives the optimal current-slot deployment, accounting for
service quality, unmet demand, battery impact, and switching.

The main contributions are as follows:
\begin{itemize}
    \item We develop a joint compression and deployment framework
    for satellite MoE-VLM services. A set of evaluated model profiles connects offline compression
    to online profile selection and distributed deployment.

    \item We design a rank allocation method based on output
    perturbation for inference under limited resources. A forward-only,
    routing-aware probe guides heterogeneous expert compression
    at specified layer-wise parameter retention ratios, improving accuracy retention
    when the parameter retention ratio is low.

    \item We formulate resource-adaptive deployment as a sequence of
    per-slot optimization problems and develop a solver based
    on profile enumeration and minimum-cost shard assignment.
    It achieves the per-slot optimum and supports fast updates
    under changing energy and communication conditions.

\item Experiments with Qwen3-VL-30B-A3B-Instruct show that
SCORAS-MoE achieves a mean accuracy of $69.9\%$ at $\beta=0.3$,
compared with $66.2\%$ for uniform rank allocation. With the common
$\beta=0.4$ profile, the deployment solver improves high-load throughput
by $5.3\%$ over Greedy and substantially reduces switching and battery
impact. It also achieves better scheduling outcomes in less time than
the evaluated proximal policy optimization (PPO) and evolutionary
baselines. Adaptive profile selection then demonstrates the benefit of connecting
compression to deployment.
\end{itemize}
\section{Related Work}
\label{sec:related_work}

The two resource challenges motivate two lines of related work:
MoE compression at low parameter retention ratios, and distributed inference
and adaptive scheduling under changing satellite conditions.

\subsection{MoE Compression and Compression Sensitivity}

Mixture-of-experts (MoE) compression exploits the substantial
redundancy and heterogeneous behavior of individual experts. Existing
methods apply quantization, pruning, expert selection, and low-rank
approximation at different granularities. EAC-MoE combines
expert-selection calibration with frequency-based pruning
\cite{chen2025eacmoe}, while automated fine-grained quantization assigns
different configurations to heterogeneous expert channels
\cite{xie2025automated}. MC-MoE uses expert- and token-level significance
to guide training-free quantization and dynamic pruning
\cite{huang2025mcmoe}. MoE-SVD instead exploits the low-rank structure
and cross-expert redundancy of expert matrices
\cite{li2025moesvd}. Together, these studies motivate heterogeneous
compression because experts and expert components do not respond
uniformly to a common compression configuration.

Allocating ranks across experts for a given target parameter count requires an
expert-level allocation signal. Singular-value-based criteria
characterize matrix compressibility by measuring how much spectral
energy is concentrated in the leading singular
values~\cite{gao2024adaptive,li2025moesvd}, whereas routing statistics
describe how frequently an expert participates in inference. Routing
analysis has revealed substantial specialization among
experts~\cite{muennighoff2025olmoe}, and expert-usage statistics have
been used for pruning and dynamic
execution~\cite{lu2024experts,chen2025eacmoe}. Although inexpensive,
neither signal alone directly characterizes the functional distortion
induced by a particular low-rank approximation.

Output-space sensitivity has also been explored in MoE compression.
Prior expert-pruning methods evaluate the change in MoE-layer outputs
after removing candidate experts~\cite{lu2024experts}, while recent
mixed-precision quantization methods use compression-induced output
distortion to characterize quantization sensitivity
~\cite{duanmu2025mxmoe}. More recent work further incorporates
loss-sensitive information into expert-level mixed-precision allocation
~\cite{deng2026gemq}. These studies establish functional output
distortion as a useful signal for MoE compression. Our setting differs
in both the compression operator and system objective: we use a
forward-only, routing-aware functional probe for heterogeneous low-rank
approximation of MoE-VLM experts and construct multiple deployable
accuracy--energy profiles for resource-adaptive distributed inference.

\subsection{Distributed Inference and Satellite Service Scheduling}


Existing distributed large language model (LLM) serving systems improve resource efficiency
through inference-phase disaggregation, request migration, and
heterogeneous execution. DistServe and Splitwise separate inference
phases according to their computation and communication
characteristics~\cite{zhong2024distserve,patel2024splitwise}, while
Llumnix supports live request migration~\cite{sun2024llumnix}. For
sparsely activated MoE models, Fiddler coordinates CPU and GPU
execution when accelerator memory cannot hold the complete expert
set~\cite{kamahori2025fiddler}.

In satellite environments, distributed inference must additionally
account for orbital mobility, intermittent connectivity, and limited
onboard resources. Existing satellite-edge systems have primarily
considered task offloading, model partitioning, and satellite--ground
collaborative execution. LEOEdge jointly addresses adaptive model
deployment and inference scheduling in large
constellations~\cite{yao2025leoedge}, SLICE partitions neural-network
execution between onboard and terrestrial
resources~\cite{chen2025slice}, and learning-based approaches coordinate
offloading, computation, and communication
decisions~\cite{fan2025satellite}. These studies establish the
importance of considering orbital mobility and communication
availability when placing computation in LEO satellite networks.

Beyond task offloading, recent work addresses more complex workloads
and energy conditions. Grace supports satellite--ground VLM inference
for remote sensing~\cite{li2026grace}, while satellite-learning systems
jointly optimize model partitioning, resource allocation, and data
transmission~\cite{lei2025joint}. PHOENIX exploits sunlight-aware task
placement to reduce battery consumption~\cite{liu2024phoenix}, and
architectures for deploying large models decompose and virtualize
model components across satellite and ground resources~\cite{shi2025satellite_lam}. Service migration under orbital
mobility has also been investigated~\cite{wu2024migration}. These approaches capture important communication, computation, energy,
and mobility constraints, but generally focus on task-level offloading
or the placement of individual model components without jointly
modeling end-to-end inference pipelines whose shards use the same
model profile.

Learning-based policies have been applied to rolling-horizon satellite
task scheduling, latency--energy-aware offloading, and
Earth-observation mission scheduling
\cite{li2025rhmappo,zhou2025satelliteppo,yao2025snnppo}. Such methods
show that adaptive policies can respond to time-varying satellite
states. Nevertheless, their actions and objectives remain centered on
tasks, missions, or offloading decisions rather than the composition
and placement of complete profile-specific inference pipelines.

SCORAS-MoE connects these two lines of work through a set of
compressed MoE-VLM profiles. Output perturbation guides offline rank
allocation, and each evaluated profile supplies an accuracy--energy
option for deployment. Online, profile composition and shard placement
are optimized together through per-slot minimum-cost assignment,
supporting fast adaptation to changing energy and communication states.

\section{System Model and Problem Formulation}
\label{sec:system_model}

\begin{table}[!hb]
\centering
\caption{Summary of key notation.}
\label{tab:key-notation}
\footnotesize
\setlength{\tabcolsep}{2pt}
\renewcommand{\arraystretch}{1.04}
\begin{tabular}{@{}p{0.39\columnwidth}p{0.55\columnwidth}@{}}
\toprule
\textbf{Notation} & \textbf{Definition} \\
\midrule
$\mathcal N,N,i$ & Satellite set, number, and index \\
$\mathcal T,T,t,\tau$ & Slot set, number, index, and duration \\
$\mathcal B,M,b$ & Profile set, number, and index \\
$\mathcal K,K,k$ & Shard set, number, and index \\
$\mathcal L,L,m$, $\mathcal E_m,e$ & MoE-layer and expert sets and indices \\
$d[t]$ & Requested number of complete pipelines \\
$\chi_i[t],\mathcal N^\chi[t]$ & Feasibility indicator and feasible satellites \\
$\mathcal G_b=(\beta_b,\alpha_b,\varepsilon_b)$ & Parameter retention ratio, accuracy, and per-shard energy profile \\
$x_{i,k,b}[t],a_i[t]$ & Shard placement and satellite service state \\
$r_b[t]$ & Number of complete pipelines using profile $b$ \\
$E_i[t],E_i^{\min},E_i^{\max}$, $L_i[t]$ & Battery state and bounds, and illumination \\
$Q[t],u[t],D[t],S[t]$ & Service quality, unmet demand, battery impact, and switches \\
$U[t]$, $\omega_q,\lambda_u,\lambda_d,\lambda_s$ & Per-slot utility and weights \\
$s^{(b)}_{m,e},r^{(b)}_{m,e}$ & Functional sensitivity score and allocated expert rank \\
$\mathbf r,\mathcal R[t]$ & Profile counts and feasible compositions \\
$\mathcal J(\mathbf r)$, $y_{i,j}[t],c_{i,k,b}[t]$ & Required instances, assignment, and edge cost \\
\bottomrule
\end{tabular}
\end{table}

Table~\ref{tab:key-notation} summarizes the principal notation used
throughout the paper.

\subsection{SCORAS-MoE Framework}

SCORAS-MoE addresses limited resources through offline compression
and distributed deployment across satellites. Online scheduling adapts
this deployment to dynamic resources. The offline stage
measures the output
perturbation of each expert, allocates ranks at several parameter
retention ratios, and evaluates the resulting complete models. Their measured
accuracy and inference energy define the profiles available to the
online stage. At the beginning of each slot, the scheduler observes
the satellite states and demand, chooses a profile composition,
and assigns the required model shards to satellites. The chosen
deployment serves the slot workload and determines the battery
and placement states passed to the next slot.

\subsection{Satellite Network and Resource Dynamics}

\begin{figure*}[t]
    \centering
    \includegraphics[width=0.96\textwidth]{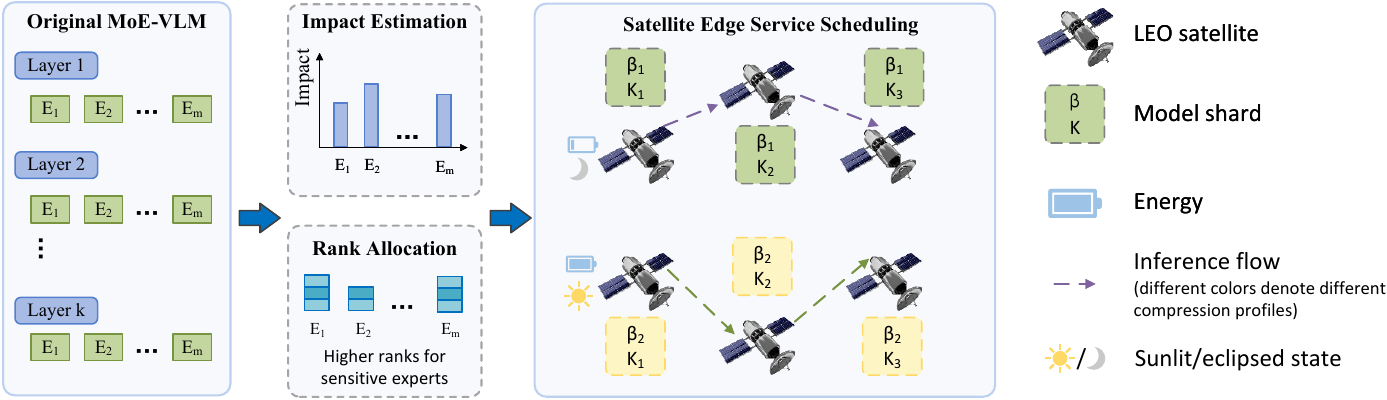}
    \caption{System setting of a Walker-Delta LEO satellite
    constellation supporting cooperative on-orbit inference.
    Each satellite is equipped with onboard computing and storage
    resources, satellite--ground and inter-satellite communication
    interfaces, a photovoltaic array, and a rechargeable battery.}
    \label{fig:system-overview}
\end{figure*}

We study the deployment of MoE-VLMs in
LEO constellations to provide inference services for both onboard
sensing applications and ground terminals through satellite--ground
links. Since the resource demand of a complete model may exceed the
compute, memory, and energy capacity of an individual satellite,
inference pipelines are partitioned and executed cooperatively across
the constellation. We consider a Walker-Delta LEO constellation
comprising $N$ satellites, indexed by
$\mathcal N=\{1,\ldots,N\}$. The satellites are distributed over
multiple orbital planes and follow predetermined near-circular
orbits. Each satellite carries an onboard computing platform and a
rechargeable battery. As illustrated in
Fig.~\ref{fig:system-overview}, processing requests within the
constellation reduces reliance on continuous access to terrestrial
data centers.

Orbital motion changes the relative satellite geometry, ground-station
visibility, inter-satellite link (ISL) connectivity, and illumination
conditions. We divide the operating horizon into equal-length scheduling slots
$\mathcal T=\{1,\ldots,T\}$, where $\tau$ denotes the duration of each
slot, following the time-slotted abstraction commonly used for dynamic
LEO edge computing~\cite{liu2024phoenix}. Satellite positions, link capacities,
illumination, and battery states are treated as constant within a slot.
Between consecutive slots, satellite positions are updated through orbital
propagation, link capacities and illumination are recalculated from the
updated geometry, and battery energy evolves according to
\eqref{eq:sm_energy}.
Throughout the paper, square brackets denote discrete-time indices.
Let $d[t]$ denote the requested number of concurrent complete inference pipelines.
Let $C_i^{\mathrm{SG}}[t]$ and $C_i^{\mathrm{ISL}}[t]$ denote the
available satellite--ground (SG) capacity and aggregate ISL capacity of satellite $i$ in
slot $t$, respectively. The corresponding bandwidth requirements per
concurrent pipeline are $\rho^{\mathrm{SG}}$ and
$\rho^{\mathrm{ISL}}$. Following capacity-constrained service placement
in satellite edge networks~\cite{inter_satellite_link}, we define the
communication-feasibility indicator as
\begin{equation}
\chi_i[t]
=
\mathbb{I}\!\left[
C_i^{\mathrm{SG}}[t]
\geq
\rho^{\mathrm{SG}}d[t]
\right]
\mathbb{I}\!\left[
C_i^{\mathrm{ISL}}[t]
\geq
\rho^{\mathrm{ISL}}d[t]
\right].
\label{eq:sm_comm}
\end{equation}
The set of satellites that meet both link requirements is
\begin{equation}
\mathcal N^{\chi}[t]
=
\left\{
i\in\mathcal N
\mid
\chi_i[t]=1
\right\}.
\label{eq:comm_feasible_set}
\end{equation}
Only satellites in $\mathcal N^{\chi}[t]$ may host an inference shard
in slot $t$.

\subsection{Sharded Pipeline Inference}

Satellite resources vary across scheduling slots. To
support resource-adaptive deployment, we compress the MoE-VLM offline
at multiple expert-projection parameter retention ratios, producing a finite
set of complete model profiles $\mathcal B$. Each profile is compressed
and evaluated independently and has a distinct parameter count,
accuracy, and inference-energy cost. This profile-based abstraction is
consistent with inference serving systems that adapt model accuracy
to available resources
~\cite{ahmad2024loki}. The compression procedure of SCORAS-MoE is
detailed in Section~\ref{sec:impact-compression}.

For distributed deployment, we partition the compressed model
corresponding to each profile at the MoE-layer granularity. Each resulting model therefore
contains $K$ required shards, indexed by
$\mathcal K=\{1,\ldots,K\}$, whose replicas can be distributed across
the constellation. Profile $b\in\mathcal B$ is represented by
\begin{equation}
\mathcal G_b
=
\left(
\beta_b,
\alpha_b,
\varepsilon_b
\right),
\label{eq:profile_interface}
\end{equation}
where $\beta_b$ is the target parameter retention ratio of expert projections, $\alpha_b$ is the measured mean accuracy
of the model compressed at retention ratio $\beta_b$, and $\varepsilon_b$ is
the measured slot-level inference energy of a representative shard under
profile $b$. Because all shards contain the same number of MoE layers
with almost identical tensor dimensions, we use $\varepsilon_b$ as the common
per-shard energy estimate for profile $b$.

A request assigned to profile $b$ is processed sequentially by one
replica of each of its $K$ shards. These replicas form a complete
inference pipeline, with all shards using the same profile. We
provision active replicas as complete pipelines, assigning one replica
of every required shard to each pipeline. Each active pipeline
processes a fixed workload of $F$ batched forward passes per slot.

Let $x_{i,k,b}[t]\in\{0,1\}$ indicate that satellite $i$ hosts shard
$k$ from profile $b$ in slot $t$. Each communication-feasible satellite can host at most one
active shard instance:
\begin{equation}
\begin{gathered}
\sum_{b\in\mathcal B}
\sum_{k\in\mathcal K}
x_{i,k,b}[t]
\leq 1,
\qquad
\forall i\in\mathcal N,
\\
x_{i,k,b}[t]
\leq \chi_i[t],
\qquad
\forall i\in\mathcal N,\;
k\in\mathcal K,\;
b\in\mathcal B.
\end{gathered}
\label{eq:sm_placement}
\end{equation}
Let $r_b[t]$ denote the number of complete pipelines using profile
$b$. Complete-pipeline provisioning requires equal active replica
counts across its shards:
\begin{equation}
\sum_{i\in\mathcal N}x_{i,k,b}[t]
=r_b[t],
\qquad
\forall k\in\mathcal K,\ b\in\mathcal B.
\label{eq:pipeline-count}
\end{equation}
Therefore, the total number of complete pipelines in slot $t$ is
$\sum_{b\in\mathcal B}r_b[t]$, corresponding to a capacity of
$F\sum_{b\in\mathcal B}r_b[t]$ batched forward passes per slot.

The service state of satellite $i$ is denoted by $a_i[t]$, where $a_i[t]=0$ means that the satellite is idle and
$a_i[t]=(k,b)$ means that it hosts shard $k$ under profile $b$. A
service switch occurs whenever $a_i[t]\neq a_i[t-1]$. This definition
covers activation, deactivation, shard replacement, and profile
replacement. Such transitions may
incur overheads such as model transfer, state synchronization, or
runtime reconfiguration~\cite{wu2024migration,sun2024llumnix}.

\subsection{Energy Model}

Satellite battery evolution is jointly determined by solar energy
harvested along the orbit and energy consumed by the inference service.
The inference load depends on whether a satellite is idle or active and,
if active, on the hosted shard. We represent this interaction using a
discrete energy-balance model for the onboard inference subsystem,
following standard charging and discharging constraints
~\cite{elgersma2024storage,razmi2026satellitebattery,
yen1993euve,miller2018space}.

For satellite $i$ in slot $t$, let $E_i[t]$ denote the battery
energy available to its onboard inference subsystem at the beginning
of the slot, and let $L_i[t]\in\{0,1\}$ indicate whether the satellite
is illuminated. The parameters $E_i^{\max}$ and
$P_i^{\mathrm{solar}}[t]$ specify the battery capacity and harvested
solar power allocated to this subsystem. Let $P_i^{\mathrm{idle}}$
denote its idle power. Given the
current placement, the profile-dependent inference energy assigned to
satellite $i$ is
\begin{equation}
\varepsilon_i^{\mathrm{inf}}[t]
=
\sum_{b\in\mathcal B}
\sum_{k\in\mathcal K}
\varepsilon_b x_{i,k,b}[t].
\label{eq:sm_inference_energy}
\end{equation}
According to \eqref{eq:sm_placement}, at most one term in this sum is nonzero. The
subsystem load and allocated solar energy in slot $t$ are
\begin{equation}
\begin{gathered}
\varepsilon_i^{\mathrm{load}}[t]
=
P_i^{\mathrm{idle}}\tau
+
\varepsilon_i^{\mathrm{inf}}[t],
\\
\varepsilon_i^{\mathrm{solar}}[t]
=
L_i[t]P_i^{\mathrm{solar}}[t]\tau.
\end{gathered}
\label{eq:sm_slot_energy}
\end{equation}
The resulting energy surplus and deficit are
\begin{equation}
\begin{gathered}
\varepsilon_i^{+}[t]
=
\left[
\varepsilon_i^{\mathrm{solar}}[t]
-
\varepsilon_i^{\mathrm{load}}[t]
\right]^+,
\\
\varepsilon_i^{-}[t]
=
\left[
\varepsilon_i^{\mathrm{load}}[t]
-
\varepsilon_i^{\mathrm{solar}}[t]
\right]^+,
\end{gathered}
\label{eq:sm_energy_balance}
\end{equation}
where $[z]^+=\max\{z,0\}$. Subject to the battery capacity and maximum
charging rate, the energy supplied for charging during slot $t$ is
\begin{equation}
\varepsilon_i^{\mathrm{ch}}[t]
=
\min\!\left\{
\varepsilon_i^{+}[t],
P_i^{\mathrm{ch,max}}\tau,
\frac{E_i^{\max}-E_i[t]}{\eta_c}
\right\},
\label{eq:sm_charge_energy}
\end{equation}
where $P_i^{\mathrm{ch,max}}$ is the maximum charging power
and $\eta_c$ is the charging efficiency. The load deficit is supplied
by the battery. A placement is energy feasible only if
\begin{equation}
\varepsilon_i^{-}[t]
\leq
\min\!\left\{
P_i^{\mathrm{dis,max}}\tau,
\eta_d\left(E_i[t]-E_i^{\min}\right)
\right\},
\label{eq:sm_discharge_feasibility}
\end{equation}
where $P_i^{\mathrm{dis,max}}$ and $\eta_d$ are the maximum discharge
power and discharge efficiency, respectively. The battery state evolves
according to
\begin{equation}
\begin{aligned}
E_i[t+1]
&=
E_i[t]
+
\eta_c\varepsilon_i^{\mathrm{ch}}[t]
-
\frac{\varepsilon_i^{-}[t]}{\eta_d},
\\
E_i^{\min}
&\leq
E_i[t+1]
\leq
E_i^{\max}.
\end{aligned}
\label{eq:sm_energy}
\end{equation}
Because $\varepsilon_i^{+}[t]$ and $\varepsilon_i^{-}[t]$ cannot both
be positive, charging and discharging do not occur in the same slot.

\subsection{Joint Optimization Formulation}

Existing satellite scheduling studies typically emphasize specific
aspects of service operation, such as energy-aware task placement and service migration
\cite{liu2024phoenix,wu2024migration}.
Continuous MoE-VLM inference requires balancing service quality,
demand fulfillment, battery impact, and service switching. We therefore formulate a joint optimization problem to improve
throughput and inference accuracy while reducing inference-induced battery
impact and service switching.

\paragraph{Service quality}
The accuracy-weighted service delivered in slot $t$ is
\begin{equation}
Q[t]
=
\sum_{b\in\mathcal B}\alpha_b r_b[t],
\label{eq:service-quality}
\end{equation}
where $r_b[t]$ is the number of complete pipelines using profile $b$,
and $\alpha_b$ is its measured mean accuracy expressed as a percentage.
Thus, $Q[t]$ jointly reflects the number and accuracy of the delivered
inference pipelines.

\paragraph{Unmet demand}
The unmet pipeline demand in slot $t$ is
\begin{equation}
u[t]
=
d[t]-\sum_{b\in\mathcal B}r_b[t].
\label{eq:unmet-demand}
\end{equation}
To avoid provisioning more pipelines than required, the aggregate
pipeline capacity is constrained by
\begin{equation}
0
\leq
\sum_{b\in\mathcal B}r_b[t]
\leq
d[t],
\qquad
u[t]
\geq
0.
\label{eq:demand-bound}
\end{equation}
\paragraph{Inference-induced battery impact}
Let $E_i[t+1]$ denote the next-slot battery energy under the scheduled
inference load, and let $E_i^{(0)}[t+1]$ denote the corresponding
next-slot energy under idle operation. Both states are determined by
\eqref{eq:sm_slot_energy}--\eqref{eq:sm_energy} under the same current
battery state and solar input. The idle case excludes only active
inference energy. The normalized marginal battery impact is
\begin{equation}
D[t]
=
\sum_{i\in\mathcal N}
\frac{
\left[
E_i^{(0)}[t+1]-E_i[t+1]
\right]^+
}{E_i^{\max}}.
\label{eq:d_costs}
\end{equation}
This term accounts for both additional battery discharge and charging
opportunities forgone because of inference.

\paragraph{Service switching cost}
Using the satellite service state $a_i[t]$, the number of switches is
\begin{equation}
S[t]
=
\sum_{i\in\mathcal N}
\mathbb{I}\!\left[
a_i[t]\neq a_i[t-1]
\right].
\label{eq:sm_costs}
\end{equation}
We set $S[1]=0$ and count switches only between consecutive slots.

Accordingly, we combine these quantities into the following per-slot
utility:
\begin{equation}
U[t]
=
\omega_q Q[t]
-
\lambda_u u[t]
-
\lambda_d D[t]
-
\lambda_s S[t],
\label{eq:sm_utility}
\end{equation}
where $\omega_q$, $\lambda_u$, $\lambda_d$, and $\lambda_s$ are
nonnegative weights for service quality, unmet demand, battery impact,
and switching, respectively.

Let $\pi$ denote a causal scheduling policy that maps the state observed
at the beginning of each slot to profile-specific pipeline counts and
shard-placement decisions. The resulting long-horizon scheduling problem
is formulated as
\begin{equation}
\begin{aligned}
\max_{\pi}\quad
&
\sum_{t\in\mathcal T}U[t]
\\
\text{s.t.}\quad
&
\eqref{eq:sm_placement},\;
\eqref{eq:pipeline-count},\;
\eqref{eq:sm_discharge_feasibility},\;
\eqref{eq:sm_energy},\;
\eqref{eq:demand-bound},
\\
&
x_{i,k,b}[t]\in\{0,1\},
\qquad
r_b[t]\in\mathbb Z_{\geq0}.
\end{aligned}
\label{eq:sm_problem}
\end{equation}
Within each slot, profile composition determines accuracy and energy
demand, while shard placement determines feasibility and switching.
Across slots, executed deployments update the battery and placement
states on which subsequent decisions depend.

Even the deterministic offline version of
\eqref{eq:sm_problem} is NP-hard. To see this, consider the restricted
case with one satellite, one shard, unit charging and discharging
efficiencies, no
solar input, zero idle power, and zero battery-impact and switching
weights. Set the unmet-demand weight to zero and let the demand be one
pipeline in every slot. Selecting profile
$b$ then yields value $\omega_q\alpha_b$ and consumes
$\varepsilon_{b}$ from a finite initial energy budget, while at most one
profile can be selected per slot. Maximizing the accumulated value over
$T$ slots is the cardinality-constrained unbounded knapsack problem,
which is NP-hard. Since this restricted case is contained in
\eqref{eq:sm_problem}, the general multi-slot problem is also NP-hard.
To respond promptly to resource changes, SCORAS-MoE optimizes
deployment at the beginning of each slot using the current state.
Section~\ref{sec:solar-orchestration} develops the minimum-cost
assignment formulation and its per-slot solution.
\section{Compression Guided by Output Perturbation}
\label{sec:impact-compression}

SCORAS-MoE allocates ranks according to
routing-weighted output perturbation, assigning higher ranks to more
sensitive experts and producing model profiles at different parameter
retention ratios.

\subsection{MoE Expert Compression}

An MoE layer routes each token to a subset of expert feed-forward
networks and aggregates their outputs. Sparse activation reduces the
computation per token, while the parameters of all experts must still
be stored. We compress the expert projection matrices using low-rank
approximations.

Let $\mathcal L=\{1,\ldots,L\}$ index the MoE layers, and let
$\mathcal{E}_m$ denote the expert set of layer $m\in\mathcal L$.
For token $n$, the router selects a top-$q$ expert subset
$\mathcal{E}^{\mathrm{top}}_{m,n}\subseteq\mathcal{E}_m$, where
$|\mathcal{E}^{\mathrm{top}}_{m,n}|=q$, and assigns a normalized routing
weight $g_{m,e,n}$ to each selected expert
$e\in\mathcal{E}^{\mathrm{top}}_{m,n}$.
Each expert is implemented as a gated feed-forward network.
For expert $e$ in layer $m$, let
$\mathbf{W}^{\mathrm{gu}}_{m,e}$ and
$\mathbf{W}^{\mathrm{d}}_{m,e}$ denote the fused gate--up and
down projection matrices, respectively. Given the token state
$\mathbf{h}_{m,n}$, the expert computation and routed MoE
output are
\begin{equation}
\begin{gathered}
\begin{bmatrix}
\mathbf{a}_{m,e,n}\\
\mathbf{b}_{m,e,n}
\end{bmatrix}
=
\mathbf{W}^{\mathrm{gu}}_{m,e}
\mathbf{h}_{m,n},                                                     \\
\mathbf{f}_{m,e}(\mathbf{h}_{m,n})
=
\mathbf{W}^{\mathrm{d}}_{m,e}
\left[
\operatorname{SiLU}(\mathbf{a}_{m,e,n})
\odot
\mathbf{b}_{m,e,n}
\right],                                                              \\
\mathbf{o}_{m,n}
=
\sum_{e\in\mathcal{E}^{\mathrm{top}}_{m,n}}
g_{m,e,n}
\mathbf{f}_{m,e}(\mathbf{h}_{m,n}).
\end{gathered}
\label{eq:moe_forward}
\end{equation}
The fused gate--up projection produces the gate and up branches.
The down projection maps their gated elementwise product back
to the hidden dimension.

To reduce the storage and computational costs of these projections, a weight matrix
$\mathbf{W}\in\mathbb{R}^{d_{\mathrm{out}}\times
d_{\mathrm{in}}}$ is approximated by two low-rank factors:
\begin{equation}
\widehat{\mathbf{W}}(r)
=
\mathbf{B}_{r}\mathbf{A}_{r},
\qquad
\mathbf{B}_{r}\in\mathbb{R}^{d_{\mathrm{out}}\times r},
\quad
\mathbf{A}_{r}\in\mathbb{R}^{r\times d_{\mathrm{in}}},
\label{eq:low_rank_factorization}
\end{equation}
where $r$ is the approximation rank. The factorization reduces
the parameter count from $d_{\mathrm{out}}d_{\mathrm{in}}$ to
$r(d_{\mathrm{out}}+d_{\mathrm{in}})$.

We use activation-aware low-rank approximation
\cite{svdllm_iclr2025,svdllmv2_naacl2025}, which minimizes projection-output
reconstruction error on calibration inputs. For a weight matrix
$\mathbf{W}$ and calibration input matrix $\mathbf{Z}$, the rank-$r$
approximation is
\begin{equation}
\widehat{\mathbf{W}}(r)
=
\arg\min_{\operatorname{rank}(\mathbf{W}')\le r}
\left\|
\mathbf{Z}\mathbf{W}^{\mathsf T}
-
\mathbf{Z}{\mathbf{W}'}^{\mathsf T}
\right\|_{\mathrm F}^{2}.
\label{eq:activation_aware_lra}
\end{equation}

We compute this approximation using regularized activation whitening
followed by truncated singular value decomposition (SVD). The calibration inputs are the routed hidden states
for the gate--up projection and the nonlinear intermediate activations
for the down projection. The allocation below determines the rank used
for each expert.

\subsection{Routing-Aware Output Perturbation}

Experts differ in their contributions to the routed layer output and
their response to low-rank approximation. To measure these differences,
we probe each expert using cached calibration activations, expert
selections, and routing weights. For a target parameter retention ratio $\beta_b$, the
parameter count of the two expert projections gives the nominal rank
for layer $m$:
\begin{equation}
\bar r_{m,b}
=
\max\!\left\{
1,
\operatorname{round}\!\left(
\beta_b
\frac{3I_mH_m}{3I_m+2H_m}
\right)
\right\},
\label{eq:cmp_nominal_rank}
\end{equation}
where $H_m$ is the hidden dimension and $I_m$ is the intermediate
dimension of each expert. We use this nominal rank to probe all experts
in the layer and determine the nominal total rank for the layer. To probe expert $e$, we temporarily replace its fused gate--up
projection with the rank-$\bar r_{m,b}$ approximation and denote the
resulting output by
$\widetilde{\mathbf{f}}^{(b)}_{m,e}(\mathbf{h}_{m,n})$.
The routing-weighted output perturbation and its aggregate score are
\begin{equation}
\begin{aligned}
\Delta^{(b)}_{m,e,n}
&=
g_{m,e,n}
\left[
\mathbf{f}_{m,e}(\mathbf{h}_{m,n})
-
\widetilde{\mathbf{f}}^{(b)}_{m,e}(\mathbf{h}_{m,n})
\right],
\\
s^{(b)}_{m,e}
&=
\left(
\sum_{n:e\in\mathcal{E}^{\mathrm{top}}_{m,n}}
\left\|\Delta^{(b)}_{m,e,n}\right\|_2^2
\right)^{1/2}.
\end{aligned}
\label{eq:cmp_impact}
\end{equation}

When only expert $e$'s fused gate--up projection is approximated
and the recorded routing decisions are retained,
$\Delta^{(b)}_{m,e,n}$ equals the resulting change in the MoE-layer
output. The aggregate score $s^{(b)}_{m,e}$ measures the
routing-weighted approximation error across calibration tokens.
Experts with larger scores receive higher ranks.

\subsection{Rank Allocation and Deployment Profiles}

We normalize the perturbation scores within each layer and allocate
its total rank among experts. For layer $m$ and profile $b$, the nominal
total rank for the layer is
\begin{equation}
R_{m,b}
=
|\mathcal E_m|\bar r_{m,b},
\label{eq:layer_rank_budget}
\end{equation}
where $|\mathcal E_m|$ is the number of experts in the layer.
Each expert first receives a minimum rank, and the
remaining total rank is distributed according to the normalized
probe scores:
\begin{equation}
\begin{gathered}
\underline r_{m,b}
=
\max\left\{
1,
\operatorname{round}
\left(\eta\bar r_{m,b}\right)
\right\},                                                            \\
w^{(b)}_{m,e}
=
\frac{s^{(b)}_{m,e}+\zeta}
{\sum_{e'\in\mathcal{E}_m}
\left(s^{(b)}_{m,e'}+\zeta\right)},                                 \\
\widetilde r^{(b)}_{m,e}
=
\underline r_{m,b}
+
\left(
R_{m,b}-|\mathcal E_m|\underline r_{m,b}
\right)
w^{(b)}_{m,e},                                                        \\
r^{(b)}_{m,e}
=
\operatorname{clip}\!\left(
\operatorname{round}\!\left[
\widetilde r^{(b)}_{m,e}
\right],
1,
r_m^{\max}
\right).
\end{gathered}
\label{eq:cmp_allocation}
\end{equation}
Here, $\widetilde r^{(b)}_{m,e}$ is the unrounded allocated rank,
$\eta$ controls the expert-level rank floor, $\zeta>0$ ensures positive allocation weights, and $r_m^{\max}$ is the maximum feasible rank in layer $m$.
The allocated rank $r^{(b)}_{m,e}$ is applied to both the gate--up and
down projections, which are reconstructed using the activation-aware
approximation in \eqref{eq:activation_aware_lra}.

Let $\mathcal V$ denote the evaluation benchmark set and $A_{b,v}$
the accuracy of profile $b$ on benchmark $v$. Its mean accuracy is
\begin{equation}
\alpha_b
=
\frac{1}{|\mathcal V|}
\sum_{v\in\mathcal V} A_{b,v}.
\label{eq:cmp_quality}
\end{equation}
Each shard of profile $b$ uses the corresponding measured shard energy
$\varepsilon_b$.

Algorithm~\ref{alg:impact_aware_compression} details the offline
compression procedure. Its inputs are the original checkpoint
$\Theta$, calibration set $\mathcal D_{\mathrm c}$, target parameter
retention ratios $\{\beta_b\}_{b\in\mathcal B}$, rank-floor coefficient
$\eta$, stabilization constant $\zeta$, and evaluation set $\mathcal V$.
Calibration caches projection inputs, expert selections, and routing
weights. These cached quantities are reused across all target retention
ratios.

For each $\beta_b$, the algorithm copies the original checkpoint and
computes the nominal rank and nominal total rank for each layer.
It probes each expert by approximating its gate--up projection at the
nominal rank, then aggregates routing-weighted output differences over
calibration tokens routed to that expert to obtain a perturbation score.
After assigning a minimum rank to every expert, it distributes the
remaining total rank according to normalized scores. The resulting
ranks are rounded and clipped, and both gate--up and down projections
are reconstructed using activation-aware low-rank approximation.

Finally, the compressed model is evaluated on $\mathcal V$ to obtain
mean accuracy $\alpha_b$. We measure the inference energy
$\varepsilon_b$ of a representative shard under the fixed
per-slot workload.
This training-free procedure returns
$\{\mathcal G_b=(\beta_b,\alpha_b,\varepsilon_b)\}_{b\in\mathcal B}$
for online deployment selection.

\begin{algorithm}[t]
\caption{MoE-VLM Compression Guided by Output Perturbation}
\label{alg:impact_aware_compression}
\begin{algorithmic}[1]
\Require Original checkpoint $\Theta$, calibration set
$\mathcal D_{\mathrm c}$, set of target parameter retention ratios
$\{\beta_b\}_{b\in\mathcal B}$, rank-floor coefficient $\eta$,
stabilization constant $\zeta$, and evaluation set $\mathcal V$
\Ensure Profile set $\{\mathcal G_b\}_{b\in\mathcal B}$

\State
$\mathcal Z_{\mathrm c}
=
\operatorname{Calibrate}
\!\left(
\Theta,\mathcal D_{\mathrm c}
\right)$

\State
$\displaystyle
\widehat{\mathbf W}(\mathbf W,\mathbf Z,r)
=
\arg\min_{\operatorname{rank}(\mathbf W')\leq r}
\left\|
\mathbf Z\mathbf W^{\mathsf T}
-
\mathbf Z{\mathbf W'}^{\mathsf T}
\right\|_{\mathrm F}^{2}$

\ForAll{$b\in\mathcal B$}

    \State
    $\Theta^{(b)}
    =
    \operatorname{Copy}
    \!\left(
    \Theta
    \right)$

    \ForAll{$m\in\mathcal L$}

        \State
        $\displaystyle
        \bar r_{m,b}
        =
        \max\!\left\{
        1,
        \operatorname{round}\!\left(
        \beta_b
        \frac{3I_mH_m}{3I_m+2H_m}
        \right)
        \right\}$

        \State
        $R_{m,b}
        =
        |\mathcal E_m|\bar r_{m,b}$

        \ForAll{$e\in\mathcal E_m$}

            \State
            $\displaystyle
            \widetilde{\mathbf W}^{(\mathrm{gu},b,\mathrm{probe})}_{m,e}
            =
            \widehat{\mathbf W}
            \!\left(
            \mathbf W^{\mathrm{gu}}_{m,e},
            \mathbf Z^{\mathrm{gu}}_{m,e},
            \bar r_{m,b}
            \right)$

            \State Construct
            $\widetilde{\mathbf f}^{(b)}_{m,e}$ using
            $\widetilde{\mathbf W}^{(\mathrm{gu},b,\mathrm{probe})}_{m,e}$

            \State
            $\displaystyle
            \Delta^{(b)}_{m,e,n}
            =
            g_{m,e,n}
            \left[
            \mathbf f_{m,e}(\mathbf h_{m,n})
            -
            \widetilde{\mathbf f}^{(b)}_{m,e}(\mathbf h_{m,n})
            \right]$

            \State
            $\displaystyle
            s^{(b)}_{m,e}
            =
            \left(
            \sum_{n:e\in\mathcal E^{\mathrm{top}}_{m,n}}
            \left\|
            \Delta^{(b)}_{m,e,n}
            \right\|_2^2
            \right)^{1/2}$

        \EndFor

        \State
        $\displaystyle
        \underline r_{m,b}
        =
        \max\!\left\{
        1,
        \operatorname{round}\!\left(
        \eta\bar r_{m,b}
        \right)
        \right\}$

        \State
        $\displaystyle
        R^{\mathrm{rem}}_{m,b}
        =
        R_{m,b}-|\mathcal E_m|\underline r_{m,b}$

        \ForAll{$e\in\mathcal E_m$}

            \State
            $\displaystyle
            w^{(b)}_{m,e}
            =
            \frac{s^{(b)}_{m,e}+\zeta}
            {\sum_{e'\in\mathcal E_m}
            \left(s^{(b)}_{m,e'}+\zeta\right)}$

            \State
            $\displaystyle
            \widetilde r^{(b)}_{m,e}
            =
            \underline r_{m,b}
            +
            R^{\mathrm{rem}}_{m,b}w^{(b)}_{m,e}$

            \State
            $\displaystyle
            r^{(b)}_{m,e}
            =
            \operatorname{clip}\!\left(
            \operatorname{round}\!\left[
            \widetilde r^{(b)}_{m,e}
            \right],
            1,r_m^{\max}
            \right)$

            \ForAll{$p\in
            \{\mathrm{gu},\mathrm{d}\}$}

                \State
                $\widetilde{\mathbf W}^{(p,b)}_{m,e}
                =
                \widehat{\mathbf W}
                \!\left(
                \mathbf W^{p}_{m,e},
                \mathbf Z^{p}_{m,e},
                r^{(b)}_{m,e}
                \right)$

                \State
                $\Theta^{(b)}
                \!\left[
                \mathbf W^{p}_{m,e}
                \right]
                =
                \widetilde{\mathbf W}^{(p,b)}_{m,e}$

            \EndFor

        \EndFor

    \EndFor

    \ForAll{$v\in\mathcal V$}

        \State
        $A_{b,v}
        =
        \operatorname{Evaluate}
        \!\left(
        \Theta^{(b)},v
        \right)$

    \EndFor

    \State
    $\displaystyle
    \alpha_b
    =
    \frac{1}{|\mathcal V|}
    \sum_{v\in\mathcal V}
    A_{b,v}$

    \State
    $\varepsilon_b
    =
    \operatorname{MeasureShardEnergy}
    \!\left(
    \Theta^{(b)}
    \right)$

    \State
    $\displaystyle
    \mathcal G_b
    =
    \left(
    \beta_b,
    \alpha_b,
    \varepsilon_b
    \right)$

\EndFor

\State
\Return
$\{\mathcal G_b\}_{b\in\mathcal B}$

\end{algorithmic}
\end{algorithm}

\section{Fast Resource-Adaptive Deployment}
\label{sec:solar-orchestration}

SCORAS-MoE adapts deployment to changing satellite resources by
enumerating profile compositions and solving a minimum-cost shard
assignment problem for each composition.

\subsection{Current-Slot State and Profile Composition}

In slot $t$, the assignment candidates are all communication-feasible
satellites defined in \eqref{eq:comm_feasible_set}:
\begin{equation}
\mathcal C[t] = \mathcal N^\chi[t].
\label{eq:orc_assignment_set}
\end{equation}
Battery energy, illumination, and the preceding placement determine
edge feasibility and cost.

Let $M=|\mathcal B|$ be the number of available profiles and
$C_t=|\mathcal C[t]|$ the number of communication-feasible satellites.
The maximum number of pipelines considered in slot $t$ is
$P_t=\min\{d[t],\lfloor C_t/K\rfloor\}$.
The scheduler enumerates profile-count vectors
$\mathbf r=(r_1,\ldots,r_M)$, where $r_b$ counts the complete pipelines
using profile $b$. The candidate set is
\begin{equation}
\mathcal R[t]
=
\left\{
\mathbf r\in\mathbb Z_{\geq0}^{M}:
\sum_{b\in\mathcal B}r_b
\leq P_t
\right\}.
\label{eq:orc_profiles}
\end{equation}
Different pipelines may use different profiles, while all shards of an
individual pipeline use the same profile.

For each $\mathbf r\in\mathcal R[t]$, the required shard instances,
indexed by profile, pipeline, and shard, are
\begin{equation}
\begin{gathered}
\mathcal J(\mathbf r)
=
\left\{
(b,\ell,k):
b\in\mathcal B,\ 
1\leq\ell\leq r_b,\ 
k\in\mathcal K
\right\},                                                    \\
|\mathcal J(\mathbf r)|
=
K\sum_{b\in\mathcal B}r_b.
\end{gathered}
\label{eq:orc_instances}
\end{equation}
The index $\ell$ distinguishes pipelines using the same profile.
The capacity bound in \eqref{eq:orc_profiles} excludes vectors satisfying
\begin{equation}
|\mathcal J(\mathbf r)|>|\mathcal C[t]|.
\label{eq:orc_size_feasibility}
\end{equation}

\subsection{Minimum-Cost Shard Assignment}

For a fixed profile-count vector, the service quality and unmet demand
are determined. Deployment then minimizes the battery impact and
switching cost of assigning its required shards to satellites.
For communication-feasible satellite $i$ and profile--shard pair $(b,k)$,
the battery-state transition during inference and its feasibility are first
evaluated using \eqref{eq:sm_slot_energy}--\eqref{eq:sm_energy}.
Let $\Phi_i[t](E,\varepsilon)$ denote the resulting single-slot battery-state
transition from initial energy $E$ when the active inference energy is
$\varepsilon$, under the slot-$t$ illumination and solar input.
For an energy-feasible edge, let
\begin{equation}
\begin{gathered}
E_i^{(0)}[t+1]
=
\Phi_i[t](E_i[t],0),
\\
E_i^{(k,b)}[t+1]
=
\Phi_i[t](E_i[t],\varepsilon_b)
\end{gathered}
\label{eq:orc_battery_states}
\end{equation}
denote the idle and active next-slot battery states under the
same current energy and solar input. The additional normalized
battery impact of the assignment is
\begin{equation}
\Delta D_{i,k,b}[t]
=
\frac{
\left[
E_i^{(0)}[t+1]-E_i^{(k,b)}[t+1]
\right]^+
}{E_i^{\max}}.
\label{eq:orc_incremental_dod}
\end{equation}
An active edge is admissible only when its load deficit
satisfies \eqref{eq:sm_discharge_feasibility}.

We first introduce the constant switching cost associated with
deactivating all previously active satellites:
\begin{equation}
C^{\mathrm{sw}}_0[t]
=
\begin{cases}
0, & t=1,\\
\displaystyle
\lambda_s
\sum_{i\in\mathcal N}
\mathbb I\!\left[a_i[t-1]\neq0\right], & t>1.
\end{cases}
\label{eq:orc_switch_constant}
\end{equation}
For a candidate assignment of profile--shard pair $(b,k)$ to
satellite $i$, the corresponding switching-cost adjustment is
\begin{equation}
\delta^{\mathrm{sw}}_{i,k,b}[t]
=
\begin{cases}
-\lambda_s, & t>1\ \text{and}\ a_i[t-1]=(k,b),\\
+\lambda_s, & t>1\ \text{and}\ a_i[t-1]=0,\\
0, & \text{otherwise}.
\end{cases}
\label{eq:orc_switch_adjustment}
\end{equation}
The first case cancels the assumed deactivation cost when the complete
profile--shard state is retained. The second case charges for activating
a previously idle satellite. Assigning a different state to an already
active satellite requires no additional adjustment because its single
state transition is already included in $C^{\mathrm{sw}}_0[t]$.

The assignment-edge cost is
\begin{equation}
c_{i,k,b}[t]
=
\begin{cases}
\lambda_d\Delta D_{i,k,b}[t]
+
\delta^{\mathrm{sw}}_{i,k,b}[t],
&
\text{if feasible},
\\
M_{\infty},
&
\text{otherwise}.
\end{cases}
\label{eq:orc_edge_cost}
\end{equation}
An energy-infeasible edge is excluded by assigning a sufficiently
large cost $M_{\infty}$. Together,
$C^{\mathrm{sw}}_0[t]$ and
$\delta^{\mathrm{sw}}_{i,k,b}[t]$ exactly account for activation,
deactivation, shard replacement, and profile replacement in the
current-slot assignment.

Let $j=(b,\ell,k)\in\mathcal J(\mathbf r)$ denote a required
shard instance, and let
$y_{i,j}[t]\in\{0,1\}$ indicate whether instance $j$ is assigned to
communication-feasible satellite $i$. Its assignment cost is directly
given by $c_{i,k,b}[t]$, which is independent of the pipeline index
$\ell$.
For each profile-count vector, SCORAS-MoE solves
\begin{equation}
\begin{aligned}
\widetilde C^{\star}[t](\mathbf r)
=
\min_{\mathbf y[t]}\quad
&
\sum_{i\in\mathcal C[t]}
\sum_{\substack{j=(b,\ell,k)\\
\in\mathcal J(\mathbf r)}}
c_{i,k,b}[t]y_{i,j}[t]
\\
\text{s.t.}\quad
&
\sum_{i\in\mathcal C[t]}
y_{i,j}[t]
=
1,
&&
\forall j\in\mathcal J(\mathbf r),
\\
&
\sum_{j\in\mathcal J(\mathbf r)}
y_{i,j}[t]
\leq1,
&&
\forall i\in\mathcal C[t],
\\
&
y_{i,j}[t]\in\{0,1\}.
\end{aligned}
\label{eq:orc_assignment}
\end{equation}
The total assignment cost is
$C^{\star}[t](\mathbf r)=C^{\mathrm{sw}}_0[t]
+\widetilde C^{\star}[t](\mathbf r)$.
The first constraint assigns every required shard instance to one
satellite, and the second allows each satellite to host at most one
instance. This is a rectangular linear sum-assignment problem, which
we solve using the Hungarian algorithm.

\subsection{Profile Selection and Per-Slot Optimality}

After solving the assignment for each feasible profile-count vector,
the scheduler compares its service benefit and deployment cost:

\begin{equation}
\begin{gathered}
\widehat U[t](\mathbf r)
=
\omega_q
\sum_{b\in\mathcal B}
\alpha_b r_b
-
\lambda_u
\left(
d[t]-
\sum_{b\in\mathcal B}r_b
\right)
-
C^{\star}[t](\mathbf r),
\\
\mathbf r^{\star}[t]
=
\arg\max_{\mathbf r\in\mathcal R[t]}
\widehat U[t](\mathbf r).
\end{gathered}
\label{eq:orc_selection}
\end{equation}
Here, $\widehat U[t](\mathbf r)$ is the current-slot utility achieved
by the minimum-cost assignment for $\mathbf r$.

\paragraph{Per-slot optimality}
Every feasible placement in the complete-pipeline model corresponds to
a vector in $\mathcal R[t]$ and an assignment of its required shards.
For a fixed vector, the Hungarian algorithm minimizes the placement
cost in \eqref{eq:orc_assignment}, while the service-quality and
unmet-demand terms remain fixed. Comparing these solutions over
$\mathcal R[t]$ therefore gives the global optimum of the single-slot
objective for the observed state.

\paragraph{Computational cost}
For $M$ profiles and a maximum of $P_t$ complete pipelines,
$\mathcal R[t]$ contains $\binom{P_t+M}{M}$ candidate vectors.
Each assignment has at most $C_t$ vertices on either side and takes
at most $O(C_t^3)$ time. The total matching work is therefore bounded
by $O\!\left(\binom{P_t+M}{M}C_t^3\right)$, with a further
$O\!\left(\binom{P_t+M}{M}M\right)$ for profile enumeration and scoring.
In practice, the number of profiles is small.

Algorithm~\ref{alg:solar_moe_scheduling} details the online procedure
for slot $t$. Its inputs are demand $d[t]$, satellite battery and
illumination states $E_i[t]$ and $L_i[t]$, communication-feasibility
indicators $\chi_i[t]$, previous service states $a_i[t-1]$, and the
offline profiles $\{\mathcal G_b\}_{b\in\mathcal B}$. It constructs
the communication-feasible satellite set and determines the maximum
pipeline count from demand and available shard capacity.

Starting with the zero-pipeline deployment and its unmet-demand and
service-deactivation costs, the scheduler enumerates nonzero
profile-count vectors. For each vector, it creates the required shard
instances and a cost matrix combining incremental battery impact and
switching-cost adjustments. Energy-infeasible edges receive a
prohibitive cost. The Hungarian algorithm computes the minimum-cost
assignment, and solutions containing infeasible edges are discarded.
Each feasible assignment is scored using accuracy-weighted throughput,
unmet demand, and total deployment cost, including the baseline
switching cost. The scheduler retains the composition and assignment
with the highest current-slot utility.

The selected assignment is decoded into a shard placement
$\mathbf x^\star[t]$. The scheduler executes the deployment and updates
battery states $E_i[t+1]$ and service states $a_i[t]$ for the next slot.
It returns the shard placement $\mathbf x^\star[t]$ and the
profile-count vector $\mathbf r^\star[t]$.

\begin{algorithm}[t]
\caption{SCORAS-MoE Per-Slot Profile Selection and Shard Assignment}
\label{alg:solar_moe_scheduling}
\begin{algorithmic}[1]
\Require Current demand $d[t]$, satellite states
$\{E_i[t],L_i[t],\chi_i[t],a_i[t-1]\}_{i\in\mathcal N}$,
and profile set $\{\mathcal G_b\}_{b\in\mathcal B}$
\Ensure Profile composition $\mathbf r^\star[t]$ and
profile-aware shard placement $\mathbf x^\star[t]$

\State Construct $\mathcal N^\chi[t]$ using \eqref{eq:comm_feasible_set}
\State $\mathcal C[t] \gets \mathcal N^\chi[t]$
\State $P_t\gets\min\{d[t],\lfloor|\mathcal C[t]|/K\rfloor\}$
\State Set $C^{\mathrm{sw}}_0[t]$ using
\eqref{eq:orc_switch_constant}

\State
$\mathcal R[t]
=
\left\{
\mathbf r\in\mathbb Z_{\geq0}^{|\mathcal B|}:
\sum_{b\in\mathcal B}r_b\leq P_t
\right\}$

\State
$\mathbf r^\star[t]=\mathbf 0$,
$\mathbf y^\star[t]=\varnothing$,
$\mathbf x^\star[t]=\mathbf 0$,
and
$\widehat U^\star[t]=-\lambda_u d[t]-C^{\mathrm{sw}}_0[t]$

\ForAll{$\mathbf r\in\mathcal R[t]\setminus\{\mathbf 0\}$}

    \State
    $\mathcal J(\mathbf r)
    =
    \left\{
    (b,\ell,k):
    b\in\mathcal B,\,
    1\leq\ell\leq r_b,\,
    k\in\mathcal K
    \right\}$

    \If{$|\mathcal J(\mathbf r)|>|\mathcal C[t]|$}
        \State \textbf{continue}
    \EndIf

    \State
    $\mathbf C[t](\mathbf r)
    =
    \left[
    c_{i,k,b}[t]
    \right]_{
    i\in\mathcal C[t],\,
    j=(b,\ell,k)\in\mathcal J(\mathbf r)
    }$

    \State
    $\left(
    \widetilde C^\star[t](\mathbf r),
    \mathbf y^\star(\mathbf r)
    \right)
    =
    \operatorname{Hungarian}
    \!\left(
    \mathbf C[t](\mathbf r)
    \right)$

    \If{the assignment contains an energy-infeasible edge}
        \State \textbf{continue}
    \EndIf

    \State
    $C^\star[t](\mathbf r)
    =
    C^{\mathrm{sw}}_0[t]
    +
    \widetilde C^\star[t](\mathbf r)$

    \State
    $\displaystyle
    \widehat U[t](\mathbf r)
    =
    \omega_q
    \sum_{b\in\mathcal B}\alpha_b r_b
    -
    \lambda_u
    \left(
    d[t]-\sum_{b\in\mathcal B}r_b
    \right)
    -
    C^\star[t](\mathbf r)$

    \If{$\widehat U[t](\mathbf r)>\widehat U^\star[t]$}
        \State
        $\widehat U^\star[t]
        =
        \widehat U[t](\mathbf r)$
        \State
        $\mathbf r^\star[t]
        =
        \mathbf r$
        \State
        $\mathbf y^\star[t]
        =
        \mathbf y^\star(\mathbf r)$
    \EndIf

\EndFor

\State
$\mathbf x^\star[t]
=
\operatorname{Decode}
\!\left(
\mathbf y^\star[t]
\right)$

\State
$\left\{
E_i[t+1],a_i[t]
\right\}_{i\in\mathcal N}
=
\operatorname{ExecuteAndUpdate}
\!\left(
\mathbf r^\star[t],
\mathbf x^\star[t]
\right)$

\State
\Return
$\mathbf r^\star[t],\mathbf x^\star[t]$

\end{algorithmic}
\end{algorithm}

\section{Performance Evaluation}
\label{sec:evaluation}

We evaluate compression accuracy, fixed-profile deployment performance
and decision time, adaptive profile composition, and deployment
ablations to assess how SCORAS-MoE supports inference under limited
and time-varying satellite resources.

\subsection{Experimental Setup}
\label{subsec:eval-setup}

\subsubsection{Model and Compression Configuration}
The model contains $L=48$ MoE layers and is divided into $K=8$
balanced shards, each containing $L/K=6$ consecutive MoE layers.
We evaluate profiles at four layer-wise parameter retention ratios,
\[
\{\beta_b\mid b\in\mathcal B\}=\{0.3,0.4,0.6,0.8\}.
\]
We compare three reference allocation methods:
Uniform~\cite{gao2024adaptive} assigns the same
nominal rank to all experts, Energy~\cite{li2025moesvd} allocates ranks
according to the retained spectral energy of the activation-weighted
expert matrix, and Frequency~\cite{chen2025eacmoe,li2024merge,huang2025mcmoe} allocates ranks
according to expert activation counts during calibration.
SCORAS-MoE allocates ranks based on the routing-weighted output
perturbation measured at the nominal rank.
All four methods share the calibration inputs, activation-aware low-rank
factorization, nominal total rank for each layer, minimum rank, and procedure
for evaluating the full model. The comparison therefore measures how the rank
allocation signal affects accuracy at each parameter retention ratio. Accuracy is measured on the TextVQA and DocVQA validation sets,
the AI2D test set, OCRBench, and RealWorldQA. The uncompressed model
obtains scores of $84.7\%$, $93.9\%$, $82.8\%$, $68.4\%$, and $73.6\%$,
respectively, with a mean of $80.7\%$.

\subsubsection{Satellite and Communication Configuration}

We construct a Walker-Delta LEO constellation with $N=60$ satellites and
generate synthetic two-line element sets (TLEs). Satellite positions are
propagated using the Simplified General Perturbations 4 (SGP4) model, and illumination is determined using the cylindrical
Earth-shadow model. The scheduling horizon contains $T=24$ slots of $300$~s.
We vary the number of requested pipelines over $n_{\mathrm{req}}\in\{5,6,7,8,9\}$.
For each load setting, $d[t]=n_{\mathrm{req}}$ for all $t\in\mathcal T$.


The four SCORAS-MoE profiles have mean accuracies of $69.9\%$, $76.1\%$,
$80.7\%$, and $82.7\%$ for $\beta=0.3$, $0.4$, $0.6$, and $0.8$,
respectively. The online scheduler uses the corresponding scores on
the $0$--$100$ scale as $\alpha_b$. We measure the dynamic inference
energy of one representative shard at each parameter retention ratio on an
NVIDIA GeForce RTX 4060 Laptop GPU. The resulting per-shard slot energies
are $2662.8$, $3146.0$, $3922.8$, and $4976.8$~J for $\beta=0.3$, $0.4$,
$0.6$, and $0.8$, respectively, and the corresponding value is applied
to every shard of the same profile.

Each active shard executes $F=400$ forward passes per slot, and each modeled
onboard inference subsystem consumes $2$~W of idle power. For each
satellite, we allocate an $8$~Wh battery-energy budget and up to $25$~W
of harvested solar power to the onboard inference subsystem when the satellite
is illuminated. These values represent the portions of the spacecraft
battery capacity and solar-array output made available to inference,
rather than the capacity and generation of the complete satellite.
Communication is enforced through the hard satellite--ground and
inter-satellite link feasibility constraints. No separate communication
power is added to the inference-subsystem energy accounting. We set
$\omega_q=10$, $\lambda_u=300$, $\lambda_d=30$, and $\lambda_s=0.01$.

\subsubsection{Baselines and Metrics}

For the deployment comparison, all methods use the same SCORAS-MoE profile
with fixed $\beta=0.4$. We compare SCORAS-MoE against three baselines.
Greedy sequentially assigns shards to communication-feasible satellites by
prioritizing sunlight and residual battery energy. The $(1+\lambda)$
evolutionary algorithm (EA) is initialized from the Greedy placement and uses
standard parent--offspring selection \cite{kazakovtsev2023pmedian}.
PPO is initialized from the same placement and implemented following the
standard framework \cite{schulman2017ppo}. EvoX is used only as the
implementation framework for vectorized evaluation of evolutionary
candidates \cite{huang2024evox}.  All methods use the same
workload traces, feasibility rules, and evaluator. 

For the joint compression--deployment experiment, SCORAS-MoE is compared with
Fixed-$\beta=0.3$, Fixed-$\beta=0.4$, Fixed-$\beta=0.6$, and
Fixed-$\beta=0.8$. These baselines use the same communication filtering,
energy model, and assignment method that accounts for switching
costs across all communication-feasible satellites. They differ in
whether deployment uses a fixed compression profile or selects its
profile composition online.

We report the number of completed pipelines, demand fulfillment ratio, inference-induced
normalized depth of discharge (DoD), unmet pipeline demand, the number
of service switches, and wall-clock solving time. For profile-selection
experiments, we additionally report accuracy-weighted throughput,
pipeline-weighted mean accuracy, and profile counts. We compute
accuracy-weighted throughput as
$\sum_{t\in\mathcal T}\sum_{b\in\mathcal B}(\alpha_b/100)r_b[t]$,
and pipeline-weighted mean accuracy as
$\sum_{t,b}\alpha_b r_b[t]/\sum_{t,b}r_b[t]$. A switch is
counted whenever a satellite's complete profile--shard state changes
between consecutive slots. A profile-only change, a shard-only change,
or a simultaneous profile and shard change is counted once for that
satellite.

\subsection{Compression Accuracy at Low Parameter Retention Ratios}
\label{subsec:compression-evaluation}

\begin{figure}[t]
\centering
\includegraphics[width=\columnwidth]{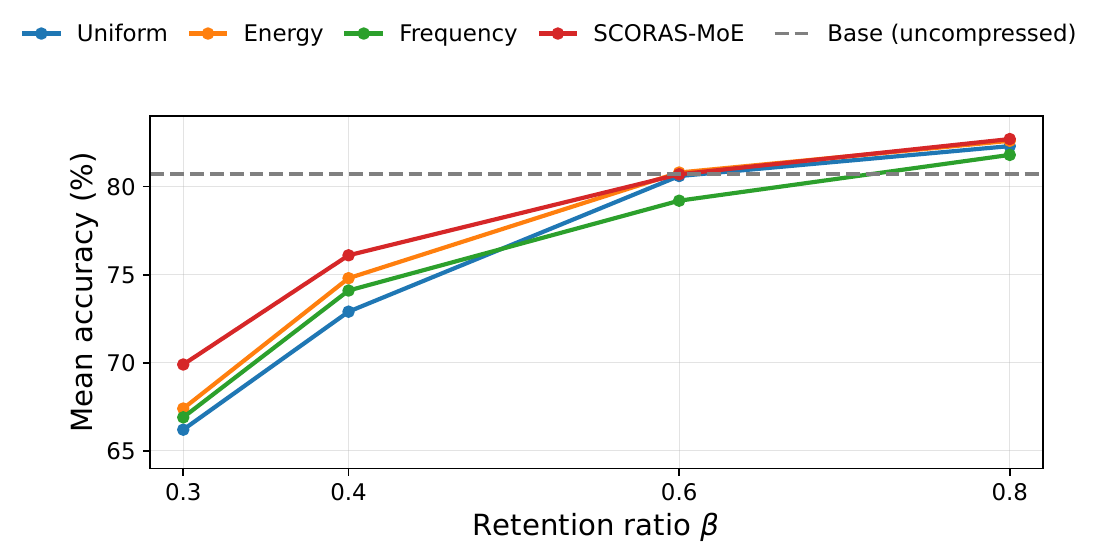}
\caption{Mean accuracy at different parameter retention ratios. The gray dashed line denotes the uncompressed base-model mean accuracy of $80.7\%$.}
\label{fig:budget-mean-accuracy}
\end{figure}

Table~\ref{tab:compression-results} and
Fig.~\ref{fig:budget-mean-accuracy} show that rank allocation based
on output perturbation is most beneficial under aggressive compression,
when the parameter retention ratio is lowest.
At $\beta=0.3$, SCORAS-MoE achieves a mean accuracy of $69.9\%$,
compared with $66.2\%$ for Uniform, $67.4\%$ for Energy,
and $66.9\%$ for Frequency.
At $\beta=0.4$, the corresponding accuracies are $76.1\%$,
$72.9\%$, $74.8\%$, and $74.1\%$, respectively.
Across the four retention ratios, SCORAS-MoE achieves the best mean score at
$\beta=0.3$, $\beta=0.4$, and $\beta=0.8$.
At $\beta=0.6$, SCORAS-MoE achieves $80.7\%$, compared with
the highest mean accuracy of $80.8\%$ achieved by Energy.

\begin{table*}[!t]
\centering
\caption{Compression accuracy (\%) of Qwen3-VL-30B-A3B-Instruct.
The best mean accuracy at each retention ratio is in bold.}
\label{tab:compression-results}
\scriptsize
\resizebox{\textwidth}{!}{%
\begin{tabular}{cccccccc}
\toprule
Retention ratio & Method & TextVQA VAL & DocVQA VAL & AI2D TEST &
OCRBench & RealWorldQA & Mean \\
\midrule
Base & Uncompressed & 84.7 & 93.9 & 82.8 & 68.4 & 73.6 & 80.7 \\
\midrule
$0.3$ & Uniform & 71.3 & 84.0 & 53.7 & 65.3 & 56.9 & 66.2 \\
$0.3$ & Energy & 73.1 & 85.6 & 54.1 & 68.4 & 55.8 & 67.4 \\
$0.3$ & Frequency & 73.4 & 84.8 & 50.1 & 67.7 & 58.4 & 66.9 \\
$0.3$ & SCORAS-MoE & 77.2 & 87.1 & 55.1 & 69.2 & 60.7 & \textbf{69.9} \\
\midrule
$0.4$ & Uniform & 78.1 & 89.1 & 65.5 & 72.0 & 60.0 & 72.9 \\
$0.4$ & Energy & 78.8 & 90.3 & 67.1 & 76.2 & 61.6 & 74.8 \\
$0.4$ & Frequency & 78.5 & 89.8 & 63.2 & 74.2 & 64.6 & 74.1 \\
$0.4$ & SCORAS-MoE & 80.4 & 91.2 & 68.3 & 76.1 & 64.4 & \textbf{76.1} \\
\midrule
$0.6$ & Uniform & 82.0 & 92.8 & 80.6 & 78.9 & 68.6 & 80.6 \\
$0.6$ & Energy & 82.5 & 93.1 & 81.1 & 79.4 & 67.8 & \textbf{80.8} \\
$0.6$ & Frequency & 81.7 & 92.8 & 74.9 & 79.0 & 67.6 & 79.2 \\
$0.6$ & SCORAS-MoE & 83.1 & 93.3 & 80.4 & 79.2 & 67.7 & 80.7 \\
\midrule
$0.8$ & Uniform & 83.6 & 93.4 & 83.2 & 79.3 & 71.8 & 82.3 \\
$0.8$ & Energy & 83.9 & 93.4 & 83.4 & 79.5 & 72.9 & 82.6 \\
$0.8$ & Frequency & 83.7 & 93.2 & 81.8 & 80.4 & 69.9 & 81.8 \\
$0.8$ & SCORAS-MoE & 83.7 & 93.6 & 84.4 & 80.6 & 71.1 & \textbf{82.7} \\
\bottomrule
\end{tabular}}
\end{table*}

The gains extend across the individual tasks. At $\beta=0.3$, SCORAS-MoE
achieves the best score on all five benchmarks. At $\beta=0.4$, it leads
on TextVQA VAL, DocVQA VAL, AI2D TEST, and the mean score, while remaining
close to the best result on OCRBench and RealWorldQA. At $\beta=0.8$,
the measured mean scores slightly exceed the uncompressed model's mean
on this benchmark suite. The larger gains at the two lower retention ratios
show the benefit of allocating ranks according to the output
perturbation caused by expert compression.

\subsection{Dynamic Deployment with Fixed $\beta=0.4$}
\label{subsec:fixed-budget-scheduling}

\begin{figure*}[!t]
\centering
\includegraphics[width=0.96\textwidth]{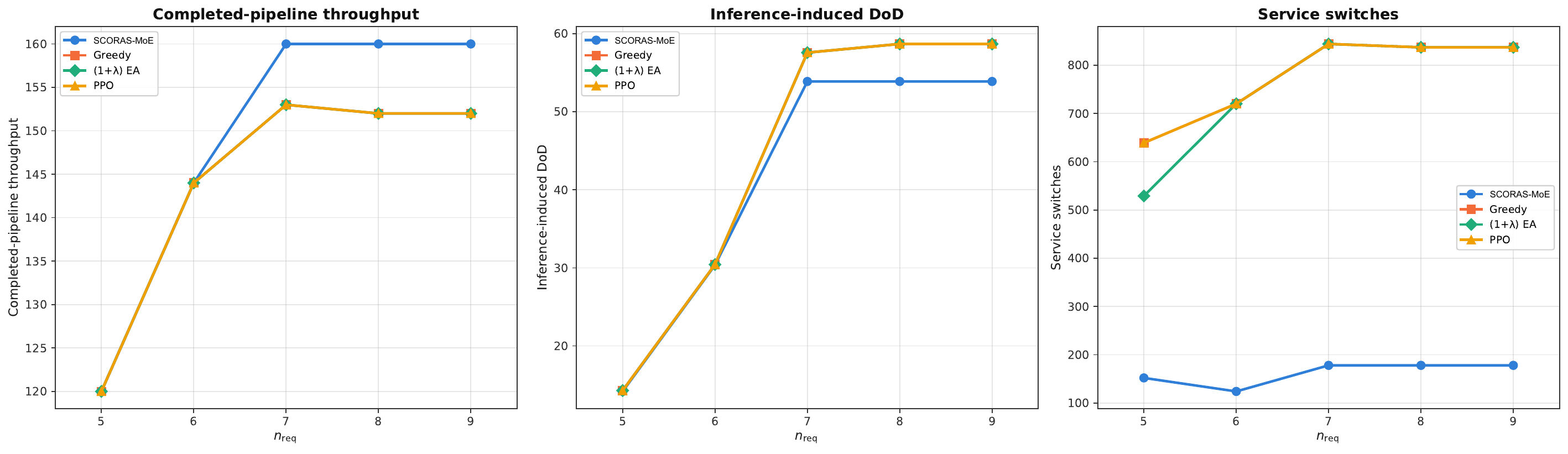}
\caption{Fixed-profile deployment at $\beta=0.4$ for
$n_{\mathrm{req}}=5$--$9$. The panels show completed pipelines,
inference-induced DoD, and service switches. All methods use the same
compressed model, workload, satellite states, and feasibility constraints.
The $(1+\lambda)$ EA uses EvoX.}
\label{fig:fixed-b04-comparison}
\end{figure*}

We evaluate deployment under changing illumination and battery states
using the same SCORAS-MoE profile at $\beta=0.4$ for all methods.

\paragraph{Deployment performance}
Figure~\ref{fig:fixed-b04-comparison} shows that SCORAS-MoE and Greedy
achieve the same throughput at $n_{\mathrm{req}}=5$ and $6$.
At $n_{\mathrm{req}}=8$ and $9$, SCORAS-MoE completes $160$ pipelines
versus $152$ for Greedy, a $5.3\%$ increase. Across the five loads,
SCORAS-MoE reduces service switches by $76.2$--$82.8\%$ relative to
Greedy. At $n_{\mathrm{req}}=9$, it also reduces inference-induced DoD
from $58.68$ to $53.88$, an $8.2\%$ reduction. The $(1+\lambda)$ EA
and PPO produce throughput and inference-induced DoD close to their
Greedy initialization across all five loads.

\paragraph{Decision time}
\label{subsec:runtime-analysis}
We measure the total wall-clock time for scheduling over all $24$ slots,
averaged over the five loads $n_{\mathrm{req}}=5$--$9$.
Greedy, fixed-profile SCORAS-MoE, the $(1+\lambda)$ evolutionary
algorithm accelerated using EvoX, and PPO take $0.006$, $0.155$,
$28.517$, and $1.352$~s, respectively. PPO training stops early when
it fails to improve the Greedy initialization by the required margin
within the first five episodes, accounting for its short runtime.
Fixed-profile SCORAS-MoE is approximately $183.5\times$ faster than
the evolutionary algorithm and $8.7\times$ faster than PPO.
Together with the service improvements, these results show that the
per-slot assignment formulation supports rapid deployment decisions
as satellite resources change.

\subsection{Joint Benefits of Compression and Adaptive Deployment}
\label{subsec:adaptive-fixed-profiles}

\begin{figure*}[!t]
\centering
\includegraphics[width=0.96\textwidth]{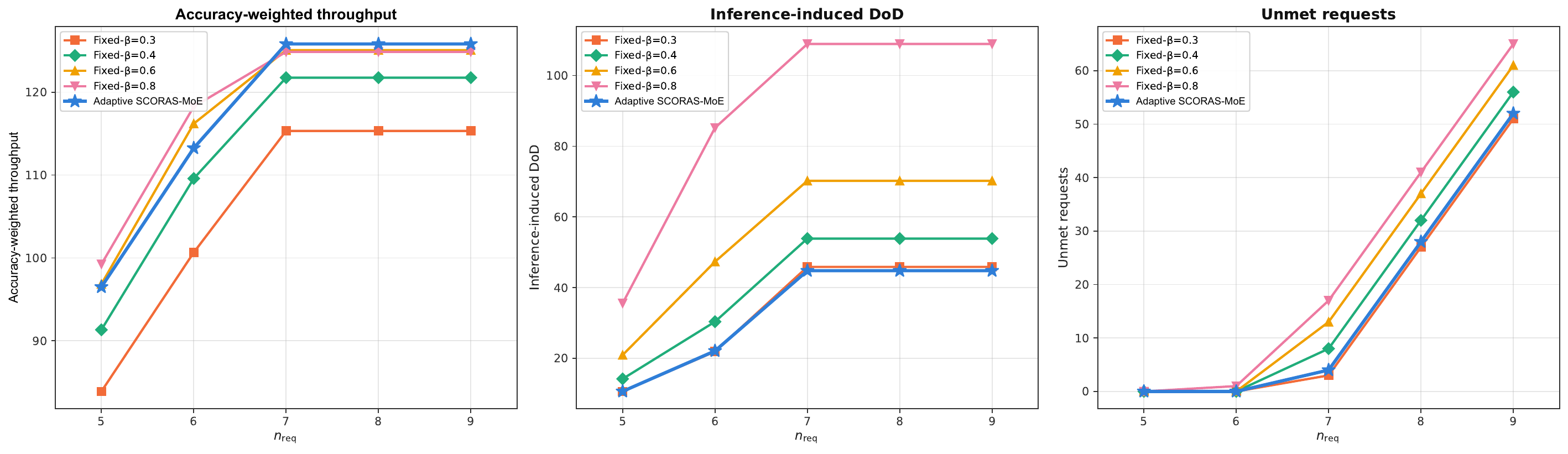}
\caption{Adaptive SCORAS-MoE profile composition versus four fixed-profile
baselines for $n_{\mathrm{req}}=5$--$9$. The panels report
accuracy-weighted throughput as defined in Section~\ref{sec:evaluation}, inference-induced DoD, and
unmet requests.}
\label{fig:adaptive-scheduling}
\end{figure*}

We compare adaptive SCORAS-MoE with four fixed-profile baselines using
the same placement solver. SCORAS-MoE selects among all four profiles
in each slot, with a common profile used by all eight shards of each
complete pipeline.

Figure~\ref{fig:adaptive-scheduling} shows that adaptive composition
maintains accuracy-weighted throughput close to the highest-quality
fixed profile while substantially reducing inference-induced DoD. At
$n_{\mathrm{req}}=7$, SCORAS-MoE completes $164$ pipelines, compared
with $151$ for Fixed-$\beta=0.8$, and reduces unmet demand from $17$
to $4$ pipelines. Its accuracy-weighted throughput increases by
$0.8\%$, while inference-induced DoD decreases by $58.9\%$.
At $n_{\mathrm{req}}=5$, it retains $97.2\%$ of the accuracy-weighted
throughput of Fixed-$\beta=0.8$ while reducing DoD by $70.1\%$.

The selected composition primarily combines the low-energy
$\beta=0.3$ profile with the high-quality $\beta=0.8$ profile.
At $n_{\mathrm{req}}=7$, these profiles account for $75$ and $86$
completed pipelines, respectively, with another $3$ using $\beta=0.4$.
This mixture sustains service quality while accommodating energy
constraints. Beyond this load, completed pipelines, accuracy-weighted
throughput, and DoD remain unchanged, while unmet demand increases
to $28$ and $52$ pipelines at $n_{\mathrm{req}}=8$ and $9$, respectively.

With adaptive profile enumeration enabled, the average total solving
time is $1.021$~s over the same $24$-slot horizon and five loads.
These results show that joint profile selection and shard placement
support fast deployment decisions.

\subsection{Ablation Study}
\label{subsec:ablation-study}

We ablate three components of SCORAS-MoE: the communication candidate
filter, the service-switching cost in the assignment, and the unmet-demand
penalty.
All variants use the measured $\beta=0.4$ profile, $K=8$ shards with six
MoE layers per shard, $F=400$ forward passes per active shard per slot, and the
same evaluator.

Table~\ref{tab:ablation-results} shows that removing the communication
candidate filter reduces the number of completed pipelines from
$120$--$160$ to $110$--$151$ and results in $10$--$65$ unmet requests.
Without this filter, all $60$ satellites enter the assignment candidate
set even when some do not satisfy the communication requirements.
Removing the service-switching cost from the assignment changes
throughput by at most one pipeline but increases service switches from
$124$--$178$ to $403$--$595$, corresponding to an increase of
$165.1$--$246.8\%$.
Removing the unmet-demand penalty causes the scheduler to conserve
battery energy at the expense of service: the number of completed pipelines
falls to $107$--$111$, while unmet demand increases to $13$--$105$.

\begin{table}[t]
\centering
\caption{Ablation results under the measured $\beta=0.4$ profile with
$F=400$ forward passes per active shard per slot. Metrics are computed over $T=24$ slots.
TP denotes the number of completed pipelines.}
\label{tab:ablation-results}
\scriptsize
\setlength{\tabcolsep}{1.2pt}
\renewcommand{\arraystretch}{0.96}
\begin{tabular*}{0.98\columnwidth}{@{\extracolsep{\fill}}clrrrrr@{}}
\toprule
$n_{\mathrm{req}}$ & Variant & TP & Fill (\%) & DoD & Unmet & Switches \\
\midrule
$5$ & Core & 120 & 100.00 & 14.18 & 0 & 152 \\
$5$ & w/o comm. candidate filter & 110 & 91.67 & 13.89 & 10 & 144 \\
$5$ & w/o switching cost & 120 & 100.00 & 14.13 & 0 & 403 \\
$5$ & w/o unmet penalty & 107 & 89.17 & 0.00 & 13 & 163 \\
\midrule
$6$ & Core & 144 & 100.00 & 30.36 & 0 & 124 \\
$6$ & w/o comm. candidate filter & 134 & 93.06 & 28.58 & 10 & 118 \\
$6$ & w/o switching cost & 144 & 100.00 & 30.33 & 0 & 430 \\
$6$ & w/o unmet penalty & 111 & 77.08 & 0.02 & 33 & 170 \\
\midrule
$7$ & Core & 160 & 95.24 & 53.88 & 8 & 178 \\
$7$ & w/o comm. candidate filter & 151 & 89.88 & 54.09 & 17 & 155 \\
$7$ & w/o switching cost & 159 & 94.64 & 52.05 & 9 & 595 \\
$7$ & w/o unmet penalty & 111 & 66.07 & 0.02 & 57 & 170 \\
\midrule
$8$ & Core & 160 & 83.33 & 53.88 & 32 & 178 \\
$8$ & w/o comm. candidate filter & 151 & 78.65 & 54.09 & 41 & 155 \\
$8$ & w/o switching cost & 159 & 82.81 & 52.05 & 33 & 595 \\
$8$ & w/o unmet penalty & 111 & 57.81 & 0.02 & 81 & 170 \\
\midrule
$9$ & Core & 160 & 74.07 & 53.88 & 56 & 178 \\
$9$ & w/o comm. candidate filter & 151 & 69.91 & 54.09 & 65 & 155 \\
$9$ & w/o switching cost & 159 & 73.61 & 52.05 & 57 & 595 \\
$9$ & w/o unmet penalty & 111 & 51.39 & 0.02 & 105 & 170 \\
\bottomrule
\end{tabular*}
\end{table}

The ablations identify how the three mechanisms support dynamic
deployment: communication filtering provides feasible assignment
candidates, the switching cost stabilizes deployment across slots,
and the unmet-demand penalty balances energy conservation with service
fulfillment.

\section{Conclusion}
\label{sec:conclusion}

Onboard VLM inference reduces raw-data downlink but must operate with
limited and dynamic resources. SCORAS-MoE combines compression with
distributed deployment to accommodate limited per-satellite capacity,
and uses online scheduling to adapt to resource changes. Offline rank
allocation based on output perturbation produces model profiles for
different parameter retention ratios. Online profile enumeration and minimum-cost assignment jointly
select the composition and shard placement, achieving the per-slot optimum.

Experiments on Qwen3-VL-30B-A3B-Instruct show that this rank allocation
is particularly effective under aggressive compression.
At $\beta=0.3$, SCORAS-MoE achieves a mean accuracy of $69.9\%$,
compared with $66.2\%$ for uniform rank allocation. The deployment solver increases throughput, reduces battery
impact and switching, and requires less computation time than the
evaluated PPO and evolutionary baselines. Adaptive profile selection
further improves the quality--energy balance over fixed-profile
deployment, demonstrating the benefit of combining compression with
fast deployment decisions.

Future work will extend the framework to heterogeneous onboard
computing platforms and latency-aware deployment in larger constellations.

\bibliographystyle{IEEEtran}
\bibliography{refs}

\begin{thebibliography}{10}
\providecommand{\url}[1]{#1}
\csname url@samestyle\endcsname
\providecommand{\newblock}{\relax}
\providecommand{\bibinfo}[2]{#2}
\providecommand{\BIBentrySTDinterwordspacing}{\spaceskip=0pt\relax}
\providecommand{\BIBentryALTinterwordstretchfactor}{4}
\providecommand{\BIBentryALTinterwordspacing}{\spaceskip=\fontdimen2\font plus
\BIBentryALTinterwordstretchfactor\fontdimen3\font minus
  \fontdimen4\font\relax}
\providecommand{\BIBforeignlanguage}[2]{{%
\expandafter\ifx\csname l@#1\endcsname\relax
\typeout{** WARNING: IEEEtran.bst: No hyphenation pattern has been}%
\typeout{** loaded for the language `#1'. Using the pattern for}%
\typeout{** the default language instead.}%
\else
\language=\csname l@#1\endcsname
\fi
#2}}
\providecommand{\BIBdecl}{\relax}
\BIBdecl

\bibitem{li2026grace}
\BIBentryALTinterwordspacing
Z.~Li, J.~Yang, Y.~Zhang, Z.~Chen, and Y.~Gao, ``Enabling near-realtime remote
  sensing via satellite–ground collaboration of large vision–language
  models,'' in \emph{Proceedings of the 2026 ACM/IEEE International Conference
  on Embedded Artificial Intelligence and Sensing Systems}, ser. SenSys
  '26.\hskip 1em plus 0.5em minus 0.4em\relax New York, NY, USA: Association
  for Computing Machinery, 2026, pp. 718--731. [Online]. Available:
  \url{https://doi.org/10.1145/3774906.3800497}
\BIBentrySTDinterwordspacing

\bibitem{delfa2026navi}
\BIBentryALTinterwordspacing
J.~M.~D. Victoria, T.~C. John, and A.~W. Herson, ``{NAVI-Orbital}: First
  in-orbit demonstration of a zero-shot vision-language model for autonomous
  {Earth} observation,'' 2026. [Online]. Available:
  \url{https://arxiv.org/abs/2606.18271}
\BIBentrySTDinterwordspacing

\bibitem{bai2025qwen3vl}
\BIBentryALTinterwordspacing
S.~Bai, Y.~Cai, R.~Chen \emph{et~al.}, ``{Qwen3-VL} technical report,'' 2025.
  [Online]. Available: \url{https://arxiv.org/abs/2511.21631}
\BIBentrySTDinterwordspacing

\bibitem{chen2025eacmoe}
\BIBentryALTinterwordspacing
Y.~Chen, Y.~Shao, P.~Wang, and J.~Cheng, ``{EAC}-{M}o{E}: Expert-selection
  aware compressor for mixture-of-experts large language models,'' in
  \emph{Proceedings of the 63rd Annual Meeting of the Association for
  Computational Linguistics (Volume 1: Long Papers)}, W.~Che, J.~Nabende,
  E.~Shutova, and M.~T. Pilehvar, Eds.\hskip 1em plus 0.5em minus 0.4em\relax
  Vienna, Austria: Association for Computational Linguistics, Jul. 2025, pp.
  12\,942--12\,963. [Online]. Available:
  \url{https://aclanthology.org/2025.acl-long.633/}
\BIBentrySTDinterwordspacing

\bibitem{xie2025automated}
\BIBentryALTinterwordspacing
Z.~Xie, Y.~Ma, X.~Zheng \emph{et~al.}, ``Automated fine-grained
  mixture-of-experts quantization,'' in \emph{Findings of the Association for
  Computational Linguistics: ACL 2025}, W.~Che, J.~Nabende, E.~Shutova, and
  M.~T. Pilehvar, Eds.\hskip 1em plus 0.5em minus 0.4em\relax Vienna, Austria:
  Association for Computational Linguistics, Jul. 2025, pp. 27\,024--27\,037.
  [Online]. Available: \url{https://aclanthology.org/2025.findings-acl.1386/}
\BIBentrySTDinterwordspacing

\bibitem{qi2026profiling}
\BIBentryALTinterwordspacing
H.~Qi, R.~Zhuo, B.~Shi \emph{et~al.}, ``Profiling-free mixed-precision
  quantization for {M}o{E} {LLM}s via fuzzy rule interpolation,'' in
  \emph{Proceedings of the 64th Annual Meeting of the {A}ssociation for
  {C}omputational {L}inguistics (Volume 1: Long Papers)}, M.~Liakata, V.~P.
  Moreira, J.~Zhang, and D.~Jurgens, Eds.\hskip 1em plus 0.5em minus
  0.4em\relax San Diego, California, United States: Association for
  Computational Linguistics, Jul. 2026, pp. 21\,484--21\,499. [Online].
  Available: \url{https://aclanthology.org/2026.acl-long.982/}
\BIBentrySTDinterwordspacing

\bibitem{lu2024experts}
\BIBentryALTinterwordspacing
X.~Lu, Q.~Liu, Y.~Xu \emph{et~al.}, ``Not all experts are equal: Efficient
  expert pruning and skipping for mixture-of-experts large language models,''
  in \emph{Proceedings of the 62nd Annual Meeting of the Association for
  Computational Linguistics (Volume 1: Long Papers)}, L.-W. Ku, A.~Martins, and
  V.~Srikumar, Eds.\hskip 1em plus 0.5em minus 0.4em\relax Bangkok, Thailand:
  Association for Computational Linguistics, Aug. 2024, pp. 6159--6172.
  [Online]. Available: \url{https://aclanthology.org/2024.acl-long.334/}
\BIBentrySTDinterwordspacing

\bibitem{duanmu2025mxmoe}
\BIBentryALTinterwordspacing
H.~Duanmu, X.~Li, Z.~Yuan \emph{et~al.}, ``{M}x{M}o{E}: Mixed-precision
  quantization for {M}o{E} with accuracy and performance co-design,'' in
  \emph{Proceedings of the 42nd International Conference on Machine Learning},
  ser. Proceedings of Machine Learning Research, A.~Singh, M.~Fazel, D.~Hsu
  \emph{et~al.}, Eds., vol. 267.\hskip 1em plus 0.5em minus 0.4em\relax PMLR,
  13--19 Jul 2025, pp. 14\,793--14\,806. [Online]. Available:
  \url{https://proceedings.mlr.press/v267/duanmu25a.html}
\BIBentrySTDinterwordspacing

\bibitem{deng2026gemq}
\BIBentryALTinterwordspacing
J.~Deng, S.~Wang, D.~Wang \emph{et~al.}, ``{GEMQ}: Global expert-level
  mixed-precision quantization for {MoE} {LLM}s,'' 2026. [Online]. Available:
  \url{https://arxiv.org/abs/2605.23078}
\BIBentrySTDinterwordspacing

\bibitem{yao2025leoedge}
S.~Yao, Y.~Lin, M.~Wang \emph{et~al.}, ``{LEOEdge}: A satellite-ground
  cooperation platform for the {AI} inference in large {LEO} constellation,''
  \emph{IEEE Journal on Selected Areas in Communications}, vol.~43, no.~1, pp.
  36--50, 2025.

\bibitem{fan2025satellite}
W.~Fan, Q.~Meng, G.~Wang, H.~Bian, Y.~Liu, and Y.~Liu, ``Satellite edge
  intelligence: {DRL}-based resource management for task inference in
  {LEO}-based satellite-ground collaborative networks,'' \emph{IEEE
  Transactions on Mobile Computing}, vol.~24, no.~10, pp. 10\,710--10\,728,
  2025.

\bibitem{chen2025slice}
Y.~Chen, Q.~Zhang, R.~Xing \emph{et~al.}, ``{SLICE}: Energy-efficient
  satellite-ground co-inference via layer-wise scheduling optimization,''
  \emph{IEEE Transactions on Services Computing}, vol.~18, no.~4, pp.
  2388--2402, 2025.

\bibitem{liu2024phoenix}
\BIBentryALTinterwordspacing
W.~Liu, Z.~Lai, Q.~Wu \emph{et~al.}, ``In-orbit processing or not?
  {S}unlight-aware task scheduling for energy-efficient space edge computing
  networks,'' in \emph{IEEE INFOCOM 2024 -- IEEE Conference on Computer
  Communications}, 2024. [Online]. Available:
  \url{https://arxiv.org/abs/2407.07337}
\BIBentrySTDinterwordspacing

\bibitem{shi2025satellite_lam}
Y.~Shi, J.~Zhu, C.~Jiang, L.~Kuang, and K.~B. Letaief, ``Satellite edge
  artificial intelligence with large models: Architectures and technologies,''
  \emph{Science China Information Sciences}, vol.~68, p. 170302, 2025.

\bibitem{huang2025mcmoe}
\BIBentryALTinterwordspacing
W.~Huang, Y.~Liao, J.~Liu \emph{et~al.}, ``Mixture compressor for
  mixture-of-experts {LLM}s gains more,'' in \emph{The Thirteenth International
  Conference on Learning Representations}, 2025. [Online]. Available:
  \url{https://openreview.net/forum?id=hheFYjOsWO}
\BIBentrySTDinterwordspacing

\bibitem{li2025moesvd}
\BIBentryALTinterwordspacing
W.~Li, L.~Li, H.~Gu \emph{et~al.}, ``{MoE-SVD}: Structured mixture-of-experts
  {LLM}s compression via singular value decomposition,'' in \emph{Proceedings
  of the 42nd International Conference on Machine Learning}, ser. Proceedings
  of Machine Learning Research, vol. 267.\hskip 1em plus 0.5em minus
  0.4em\relax PMLR, 2025, pp. 35\,209--35\,230. [Online]. Available:
  \url{https://proceedings.mlr.press/v267/li25az.html}
\BIBentrySTDinterwordspacing

\bibitem{gao2024adaptive}
\BIBentryALTinterwordspacing
S.~Gao, T.~Hua, Y.-C. Hsu, Y.~Shen, and H.~Jin, ``Adaptive rank selections for
  low-rank approximation of language models,'' in \emph{Proceedings of the 2024
  Conference of the North American Chapter of the Association for Computational
  Linguistics: Human Language Technologies (Volume 1: Long Papers)}, K.~Duh,
  H.~Gomez, and S.~Bethard, Eds.\hskip 1em plus 0.5em minus 0.4em\relax Mexico
  City, Mexico: Association for Computational Linguistics, Jun. 2024, pp.
  227--241. [Online]. Available:
  \url{https://aclanthology.org/2024.naacl-long.13/}
\BIBentrySTDinterwordspacing

\bibitem{muennighoff2025olmoe}
\BIBentryALTinterwordspacing
N.~Muennighoff, L.~Soldaini, D.~Groeneveld \emph{et~al.}, ``{OLMoE}: Open
  mixture-of-experts language models,'' 2025. [Online]. Available:
  \url{https://arxiv.org/abs/2409.02060}
\BIBentrySTDinterwordspacing

\bibitem{zhong2024distserve}
\BIBentryALTinterwordspacing
Y.~Zhong, S.~Liu, J.~Chen \emph{et~al.}, ``{DistServe}: Disaggregating prefill
  and decoding for goodput-optimized large language model serving,'' in
  \emph{18th USENIX Symposium on Operating Systems Design and Implementation
  (OSDI 24)}.\hskip 1em plus 0.5em minus 0.4em\relax Santa Clara, CA: USENIX
  Association, Jul. 2024, pp. 193--210. [Online]. Available:
  \url{https://www.usenix.org/conference/osdi24/presentation/zhong-yinmin}
\BIBentrySTDinterwordspacing

\bibitem{patel2024splitwise}
P.~Patel, E.~Choukse, C.~Zhang \emph{et~al.}, ``Splitwise: Efficient generative
  {LLM} inference using phase splitting,'' in \emph{2024 ACM/IEEE 51st Annual
  International Symposium on Computer Architecture (ISCA)}, 2024, pp. 118--132.

\bibitem{sun2024llumnix}
\BIBentryALTinterwordspacing
B.~Sun, Z.~Huang, H.~Zhao \emph{et~al.}, ``Llumnix: Dynamic scheduling for
  large language model serving,'' in \emph{18th USENIX Symposium on Operating
  Systems Design and Implementation (OSDI 24)}.\hskip 1em plus 0.5em minus
  0.4em\relax Santa Clara, CA: USENIX Association, Jul. 2024, pp. 173--191.
  [Online]. Available:
  \url{https://www.usenix.org/conference/osdi24/presentation/sun-biao}
\BIBentrySTDinterwordspacing

\bibitem{kamahori2025fiddler}
\BIBentryALTinterwordspacing
K.~Kamahori, T.~Tang, Y.~Gu, K.~Zhu, and B.~Kasikci, ``Fiddler: {CPU-GPU}
  orchestration for fast inference of mixture-of-experts models,'' in
  \emph{International Conference on Learning Representations}, Y.~Yue, A.~Garg,
  N.~Peng, F.~Sha, and R.~Yu, Eds., vol. 2025, 2025, pp. 56\,099--56\,115.
  [Online]. Available:
  \url{https://proceedings.iclr.cc/paper_files/paper/2025/file/8cd1ce03ea58b3d7dfd809e4d42f08ea-Paper-Conference.pdf}
\BIBentrySTDinterwordspacing

\bibitem{lei2025joint}
C.~Lei, S.~Wu, Y.~Yang \emph{et~al.}, ``Joint partitioning, allocation, and
  transmission optimization for federated learning in satellite constellations
  via multi-task {MARL},'' \emph{IEEE Transactions on Mobile Computing},
  vol.~24, no.~10, pp. 10\,345--10\,362, 2025.

\bibitem{wu2024migration}
H.~Wu, X.~Yang, and Z.~Bu, ``Task offloading with service migration for
  satellite edge computing: A deep reinforcement learning approach,''
  \emph{IEEE Access}, vol.~12, pp. 25\,844--25\,856, 2024.

\bibitem{li2025rhmappo}
\BIBentryALTinterwordspacing
Z.~Li, X.~Zhu, C.~Liu \emph{et~al.}, ``Dynamic task scheduling optimization by
  rolling horizon deep reinforcement learning for distributed satellite
  system,'' \emph{Expert Systems with Applications}, vol. 289, p. 128350, 2025.
  [Online]. Available:
  \url{https://www.sciencedirect.com/science/article/pii/S0957417425019694}
\BIBentrySTDinterwordspacing

\bibitem{zhou2025satelliteppo}
J.~Zhou, J.~Liang, L.~Zhao, S.~Wan, H.~Cai, and F.~Xiao, ``Latency-energy
  efficient task offloading in the satellite network-assisted edge computing
  via deep reinforcement learning,'' \emph{IEEE Transactions on Mobile
  Computing}, vol.~24, no.~4, pp. 2644--2659, 2025.

\bibitem{yao2025snnppo}
\BIBentryALTinterwordspacing
W.~Yao, X.~Shen, G.~Zhang \emph{et~al.}, ``A spiking neural network based
  proximal policy optimization method for multi-point imaging mission
  scheduling of {Earth} observation satellite,'' \emph{Swarm and Evolutionary
  Computation}, vol.~94, p. 101867, 2025. [Online]. Available:
  \url{https://www.sciencedirect.com/science/article/pii/S2210650225000252}
\BIBentrySTDinterwordspacing

\bibitem{inter_satellite_link}
H.~Hu, K.~Song, C.~Zhan, R.~Fan, and J.~Yang, ``Joint service caching and
  resource allocation over different timescales in satellite edge computing
  networks,'' \emph{IEEE Transactions on Mobile Computing}, vol.~24, no.~7, pp.
  5649--5664, 2025.

\bibitem{ahmad2024loki}
\BIBentryALTinterwordspacing
S.~Ahmad, H.~Guan, and R.~K. Sitaraman, ``Loki: A system for serving {ML}
  inference pipelines with hardware and accuracy scaling,'' in
  \emph{Proceedings of the 33rd International Symposium on High-Performance
  Parallel and Distributed Computing}, ser. HPDC '24.\hskip 1em plus 0.5em
  minus 0.4em\relax New York, NY, USA: Association for Computing Machinery,
  2024, pp. 267--280. [Online]. Available:
  \url{https://doi.org/10.1145/3625549.3658688}
\BIBentrySTDinterwordspacing

\bibitem{elgersma2024storage}
M.~B. Elgersma, G.~Morales-España, K.~I. Aardal, N.~Helistö, J.~Kiviluoma,
  and M.~M. de~Weerdt, ``Tight {MIP} formulations for optimal operation and
  investment of storage including reserves,'' \emph{IEEE Transactions on Power
  Systems}, vol.~41, no.~4, pp. 2428--2440, 2026.

\bibitem{razmi2026satellitebattery}
N.~Razmi, B.~Matthiesen, R.~Teodorescu, A.~Dekorsy, and P.~Popovski,
  ``Satellite battery lifetime extension via scheduling of onboard computing
  and energy harvesting,'' \emph{IEEE Access}, vol.~14, pp. 13\,025--13\,040,
  2026.

\bibitem{yen1993euve}
W.~L. Yen, R.~G. Littlefield, D.~R. McLean, A.~Tuchman, T.~A. Broseghini, and
  B.~J. Page, ``A battery power model for the {EUVE} spacecraft,'' in
  \emph{Proceedings of the 9th AIAA Computing in Aerospace Conference}, 1993,
  {AIAA} Paper 93-4638.

\bibitem{miller2018space}
\BIBentryALTinterwordspacing
S.~Miller, B.~T. Klefman, S.~Korn, T.~Nowden, A.~M. Delleur, and D.~McKissock,
  ``\BIBforeignlanguage{en}{The {SPACE} computer code for analyzing the
  {International Space Station} electrical power system: Past, present, and
  future},'' in \emph{\BIBforeignlanguage{en}{2018 {International} {Energy}
  {Conversion} {Engineering} {Conference}}}.\hskip 1em plus 0.5em minus
  0.4em\relax Cincinnati, Ohio: American Institute of Aeronautics and
  Astronautics, Jul. 2018. [Online]. Available:
  \url{https://arc.aiaa.org/doi/10.2514/6.2018-4635}
\BIBentrySTDinterwordspacing

\bibitem{svdllm_iclr2025}
X.~Wang, Y.~Zheng, Z.~Wan, and M.~Zhang, ``{SVD-LLM}: Truncation-aware singular
  value decomposition for large language model compression,'' in
  \emph{International Conference on Learning Representations}, 2025.

\bibitem{svdllmv2_naacl2025}
X.~Wang, S.~Alam, Z.~Wan, H.~Shen, and M.~Zhang, ``{SVD-LLM V2}: Optimizing
  singular value truncation for large language model compression,'' in
  \emph{Proceedings of the 2025 Conference of the Nations of the Americas
  Chapter of the Association for Computational Linguistics}, 2025, pp.
  4287--4296.

\bibitem{li2024merge}
\BIBentryALTinterwordspacing
P.~Li, Z.~Zhang, P.~Yadav \emph{et~al.}, ``Merge, then compress: Demystify
  efficient {SMoE} with hints from its routing policy,'' in \emph{International
  Conference on Learning Representations}, B.~Kim, Y.~Yue, S.~Chaudhuri,
  K.~Fragkiadaki, M.~Khan, and Y.~Sun, Eds., vol. 2024, 2024, pp.
  14\,234--14\,256. [Online]. Available:
  \url{https://proceedings.iclr.cc/paper_files/paper/2024/file/3d09a88c3372cdb79401fde592ca4db8-Paper-Conference.pdf}
\BIBentrySTDinterwordspacing

\bibitem{kazakovtsev2023pmedian}
L.~Kazakovtsev, I.~Rozhnov, and V.~Kazakovtsev, ``A (1+ $\lambda$) evolutionary
  algorithm with the greedy agglomerative mutation for p-median problems,'' in
  \emph{AIP Conference Proceedings}, vol. 2700, no.~1.\hskip 1em plus 0.5em
  minus 0.4em\relax AIP Publishing LLC, 2023, p. 040003.

\bibitem{schulman2017ppo}
\BIBentryALTinterwordspacing
J.~Schulman, F.~Wolski, P.~Dhariwal, A.~Radford, and O.~Klimov, ``Proximal
  policy optimization algorithms,'' 2017. [Online]. Available:
  \url{https://arxiv.org/abs/1707.06347}
\BIBentrySTDinterwordspacing

\bibitem{huang2024evox}
\BIBentryALTinterwordspacing
B.~Huang, R.~Cheng, Z.~Li, Y.~Jin, and K.~C. Tan, ``{EvoX}: A distributed
  {GPU}-accelerated framework for scalable evolutionary computation,'' 2024.
  [Online]. Available: \url{https://arxiv.org/abs/2301.12457}
\BIBentrySTDinterwordspacing

\end{thebibliography}

\end{document}